# Stochastic Frontier Analysis with Generalized Errors: inference, model comparison and averaging

This version: October 11, 2020

Kamil Makieła*  
Cracow University of Economics  
Poland  
orcid.org/0000-0003-1236-2423

Błażej Mazur  
Cracow University of Economics  
Poland  
orcid.org/0000-0001-5096-5175

## Abstract

Contribution of this paper lies in the formulation and estimation of a generalized model for stochastic frontier analysis (SFA) that nests virtually all forms used and includes some that have not been considered so far. The model is based on the generalized *t* distribution for the observation error and the generalized beta distribution of the second kind for the inefficiency-related term. We use this general error structure framework for formal testing, to compare alternative specifications and to conduct model averaging. This allows us to deal with model specification uncertainty, which is one of the main unresolved issues in SFA, and to relax a number of potentially restrictive assumptions embedded within existing SF models. We also develop Bayesian inference methods that are less restrictive compared to the ones used so far and demonstrate feasible approximate alternatives based on maximum likelihood.

**Funding:**   This research is financed by Polish National Science Center (NCN), grant number: UMO-2018/31/B/HS4/01565

* Corresponding author, Department of Econometrics and Operational Research,
 Cracow University of Economics, Rakowicka 27, 31-510 Krakow, Poland,
 email: kamil.makiela@uek.krakow.pl

# 1 Introduction

Formal research on fundamentals of productivity and productive inefficiency can be traced back to Debreu (1951) and Koopmans (1951) who provided the first formal definition of technical inefficiency as "the result of managerial choice" (Koopmans, 1951; p. 34). The first empirical technical inefficiency analysis was conducted by Farrell (1957) using data on US agriculture industry. He estimated the so-called technological frontier, also known as the best practices frontier, which describes the maximum output given inputs or alternatively, or the minimum inputs required to produce a level of output. The interest in measuring productivity and inefficiency has grown enormously since then. It encompasses numerous approaches, estimation methods and most notably an impressive diversity in fields of applications. These include, e.g., agriculture (e.g. Marzec and Pisulewski, 2017), healthcare (e.g. Koop et al., 1997), energy economics (e.g. Kumbhakar and Lien, 2017; Farsi et al., 2006), banking (e.g. Tran et al., 2020; Inanoglu et al., 2016), transportation (e.g. Stead et al., 2019), tourism (e.g., Assaf and Tsionas, 2019), education (e.g. Mayston, 2003) and even macro-level studies of entire economies (Makieła and Ouattara, 2018; Koop et al., 1999, 2000; Fare et al., 1994). The interest in frontier analysis has spread beyond just economics and management sciences; see, e.g., Fried et al. (2008, pp. 16-19) for a lengthy list of applications in about fifty different fields. This is largely due to the fact that the ability to quantify efficiency provides management with a reliable mechanism to monitor and control their performances. Furthermore, these methods have also become crucial tools for designing credible, performance-oriented policies (see, e.g., Makieła and Osiewalski, 2018; and works cited therein).

Methodologies used in the frontier analysis literature can be generally divided into two approaches. The first one is a non-parametric, mathematical programming approach dominated by Data Envelopment Analysis (DEA hereafter) proposed by Charnes et al. (1978) and its derivatives. The second one is the econometric approach developed independently by Aigner et al. (1977) and Meeusen and van den Broeck (1977); see Greene (2008), or Parmeter and Kumbhakar (2014) for an overview. This paper focuses on the econometric approach, which is currently dominated by Stochastic Frontier Analysis (SFA hereafter).

According to SFA observables in economic analyses are driven not only by some underlying structural process governing the unobserved, potential quantities, but also by random disturbances of (at least) two kinds. The first one represents observation/measurement error typical in econometric analyses, having standard properties and no tendency. The second one reflects imperfect operating mode of real economic agents and hence clearly has adverse effects on economic outcomes. It has an economic interpretation in terms of inefficiency and its statistical properties resulting from this interpretation are non-standard. This allows us to capture subtle differences in the "noise" asymmetry and provide meaningful interpretation to its sources (as inefficiency). Since uncertainty measurement is accounted for in SFA, we can relatively easily move beyond simple point estimates of inefficiency, analyze its entire distribution and formally test possible model specifications. Moreover, a single-stage approach to analyzing efficiency determinants is also available within SFA (via VED-SFA; see Koop et al., 1997).

The main goal of SFA is to estimate (in)efficiencies of decision making units and the results can vary dependently on the stochastic assumptions made on the compound error. Despite long history of development, drawbacks and limitations of existing SF methods are still present, which is crucial from the research point of view. In particular, the use of existing methods might result in an implicit introduction of strong, possibly irrelevant assumptions without empirical verification. This is very likely to produce an erroneous quantification of strengths of statistical evidence in favor of certain hypotheses regarding the technology or inefficiency processes. Consequences, such as distorted inference about structural properties of the analyzed processes or spurious detection of non-existing phenomena, are therefore likely to happen. Thus, we wish to propose an improved model framework and inference procedures that alleviate these adverse effects.

The purpose of this paper is to deal with the following four objectives, which are important for further SFA development: (1) introduction of highly flexible forms of distributions for observation errors and inefficiency terms; (2) model comparison and averaging (by encompassing key existing model specifications), (3) estimation of model parameters, and (4) inference on object-specific latent (in)efficiencies. Obviously, these objectives are interrelated: sufficient generalization of contemporary specifications provides grounds for model comparison (or averaging), while model comparison provides prerequisites for adequate treatment of model uncertainty in order to make proper inference on parameters and latent variables. Also, we emphasize that though model comparison and averaging procedures are usually used for covariate selection, we employ them here to address the problem of model specification uncertainty in SFA.

Our estimation strategy is based on the analysis of the integrated likelihood, with latent variables representing inefficiency terms integrated out numerically. We demonstrate that for the most general model, and its nested/limiting sub-cases, this process is feasible. Despite the high level of stochastic complexity, it is possible to use standard likelihood-based inference methods (ML or Bayesian). We consider two general estimation strategies: one being approximate, avoiding numerically intensive methods, and one exact (i.e. fully Bayesian), providing a formal way of dealing with all the aspects of estimation uncertainty at the cost of additional numerical complexity. For both strategies we consider model averaging via mixture pool, conceptually identical to Bayesian model averaging (BMA).

The paper is structured as follows. Section 2 provides an overview of the existing SFA literature with respect to distributional assumptions about inefficiency and observation error. Section 3 contains main distributional assumptions of the model and a discussion of nested special cases. Section 4 describes statistical inference methods. Further extensions of the basic model are discussed in Section 5. Finally, Section 6 presents the empirical examples which demonstrate feasibility of the presented methods, including estimation, model comparison and averaging. Section 7 concludes with a discussion.

## 2  Existing SFA specifications: an overview

New developments in SFA usually go in one, or several, of the following directions: (i) new distributions for inefficiency, (ii) new distributions for the random disturbance, (iii) more complex composed error for panel data modelling (i.e., adding more components; see, e.g., Colombi et al., 2014), (iv) analysis of inefficiency determinants (Koop et al., 1997; Parmeter et al., 2016), or (v) investigation of the very existence of inefficiency in the analyzed data (see, e.g., Kumbhakar et al., 2013). In this section we summarize research in the first two areas as this is the focal point of the paper. The reader should note, however, that some aspects discussed in Section 5 also relate to (iii) and (iv).

The oldest and probably the most intensely researched field in SFA concerns studying the specification of inefficiency distribution. The first proposed extension to the basic SFA models by Aigner et al. (1977) and Meeusen and van den Broeck (1977) is probably the normal-truncated-normal model by Stevenson (1980); see also van den Broeck et al. (1994), Tsionas (2002). It is a natural generalization of the normal-half-normal model, in which location of the truncation point is not necessarily at the mode of the underlying two-sided distribution. It should be noted, however, that as van den Broeck et al. (1994) point out, this model can be troublesome because of very poor identification. That is, multiple combinations of the two parameters' values that govern this distribution (location and scale) after truncation (at zero) can lead to very similar shapes. Also, if the location parameter is less than zero the resulting shape after truncation becomes similar to the exponential case. Another popular extension is the normal-gamma model, proposed by Beckers and Hammond (1987) and Greene (1990), which generalizes the normal-exponential model introduced by Meeusen and van den Broeck (1977). Although numerically challenging to estimate, see e.g. Ritter and Simar (1997), it provides a good middle ground between half-normal and exponential, which can be viewed as its two extreme cases (Greene, 2008). Other adoptions of this model have been proposed in the literature; see, e.g., van den Broeck et al. (1994) and Tsionas (2002) for normal-Erlang models. An interesting generalization is also proposed by Tancredi (2002) who uses half-Student's *t* distribution. Tsionas (2006), on the other hand, proposes the normal-lognormal model which allows us to model inefficiency over time as a stationary AR(1) dynamic process (see also Emvalomatis, 2012). This, however, comes at a cost. Lognormal distribution has virtually no probability mass near zero, which implies that full efficiency is excluded by construction. This technical assumption is restrictive and possibly the reason why this very interesting concept of dynamic inefficiency model has not received much attention in SFA. Tsionas (2007) also studies models in which inefficiency follows Weibull distribution which, he argues, is better at handling outliers. The most general distribution for inefficiency up to this date has been probably proposed by Griffin and Steel (2008) who develop an SF model based on generalized gamma distribution. Apart from its flexibility the model nests most, if not all, distributions proposed for inefficiency. Furthermore, some contemporary research has turned its attention to Rayleigh distribution, which is a special case of Weibull (Hajargasht, 2015) or experimented with a uniform and half-Cauchy distributions (Nguyen, 2010). Horrace and Parmeter (2018) also consider truncated-Laplace distribution for inefficiency, though only its simplified, exponential case is used in application. Finally, Parmeter et al. (2019) and Isaksson et al. (2020) discuss some averaging techniques for weighting conditional inefficiency. These estimators, however, are applicable to a limited class of SF models. This large volume of proposals for inefficiency distribution comes from the fact that the theory provides very limited information as to how inefficiency distribution should look like. Oikawa (2016) provides some evidence in favor of gamma distribution. This, however, is dismissed by Tsionas (2017) who argues that inefficiency does not follow any of the distributions known so far. Hence, further research is required to pursue more flexible distributions that can better reflect inefficiency.

The random symmetric disturbance (i.e., observation error) is traditionally assumed to be normally distributed in SFA. This, however, may be too restrictive in practice especially in the presence of outliers or sample heterogeneity (not only in SFA; see, e.g., Lange et al., 1989). That is why more attention has been given recently towards relaxing assumptions about the observation error. Tancredi (2002) proposes to use Student's $t$ distribution and develops the Student's $t$-half-Student's $t$ model, which directly generalizes the normal-half-normal case. Later, Tchumtchoua and Dey (2007) estimate this model using Bayesian inference and Griffin and Steel (2007) briefly discuss Bayesian estimation of Student's $t$-half-normal, Student's $t$-exponential and Student's $t$-gamma SF models using WinBUGS software package. A good overview of contemporary methods for handling outliers is provided by Stead et al. (2018) who propose a logistic-half-normal model and Stead et al. (2019) who use Student's $t$ distribution. Along the same line Wheat et al. (2019) use a mixture of normal distributions. It should be noted, however, that though a two-component mixture of normally distributed variables is indeed more flexible, from the formal viewpoint its tails are still Gaussian (hence not heavy). Horrace and Wang (2020) have recently proposed non-parametric tests to check if one deals with heavy tails in the observation error term (or inefficiency).

To sum up, all of the abovementioned proposals are interesting. However, there is no SF model general enough to encompass all of the relevant specifications which could provide a basis for model specification search, comparison or even inference pooling (i.e., model averaging). Existing SF models and inference methods are formulated using rather strong assumptions and only rarely verified. These assumptions are motivated mainly by computational or inferential convenience and are therefore likely to be empirically implausible in some cases. Validity of inference methods relies in turn upon adequacy of underlying stochastic assumptions. Consequently, methods that fail to account for specific, non-typical random properties of actual observables might lead to inaccurate structural inference on determinants influencing the potential quantity as well as identification of the factors driving inefficiency. Moreover, in many cases existing inference methods provide results that may be misinterpreted because of failure to account for alternative formulations that have practically the same explanatory power. The problem is closely related to that of weak identification; see "near-identifiability" issue in Bandyopadhyay and Das (2008).

## 3 Formulation of the GT-GB2 model

Consider the following observation equation arising from a multiplicative formulation $y_t = g(\mathbf{x}_t; \boldsymbol{\beta}) r_t \zeta_t$ (see, e.g., Koop and Steel, 2007; eq. 3):

$$\ln y_t = \ln g(\mathbf{x}_t; \boldsymbol{\beta}) - \omega \mathrm{u}_t + \mathrm{v}_t, \quad t = 1, \dots, T \tag{1}$$

with $\mathrm{v}_t$ representing the *i.i.d.* error term with symmetric and unimodal pdf, $\mathrm{u}_t$ being an *i.i.d.* latent variable, taking strictly positive values, representing the inefficiency term, and with known $\omega \in \{-1,1\}$; thus the compound error is $\varepsilon_t = \mathrm{v}_t - \omega \mathrm{u}_t$. We assume that $g(\mathbf{x}; \boldsymbol{\beta})$ is a known function (of unknown parameters $\boldsymbol{\beta}$ and exogenous variables $\mathbf{x}$) representing the frontier under consideration; it corresponds to, e.g., cost function if $\omega = -1$, or production function if $\omega = 1$, while $y_{it}$ denotes observable cost or output. Moreover, we make the following assumptions about $\mathrm{v}_t$'s and $\mathrm{u}_t$'s:

**A1.** $\mathrm{v}_t$'s are *i.i.d.* variables, with zero median, symmetric, unimodal and continuous pdf's: $p_\mathrm{v}(\mathrm{v}; \boldsymbol{\theta}^{(\mathrm{v})})$, $\mathrm{v} \in \mathbb{R}$;

**A2.** $\mathrm{u}_t$'s are nonnegative *i.i.d.* variables, having continuous pdf's: $p_\mathrm{u}(\mathrm{u}; \boldsymbol{\theta}^{(\mathrm{u})})$, $\mathrm{u} \in \mathbb{R}_+$;

**A3.** all $\mathrm{v}_t$'s are stochastically independent from all $\mathrm{u}_t$'s.

In our view, the assumptions A1-A3 are minimalistic and motivated mostly by considerations regarding statistical identification of parameters. In particular, we argue that given sufficiently general form of $p_\mathrm{u}(.)$, and $p_\mathrm{v}(.)$, the potential gain in terms of statistical fit arising from the relaxation of A1-A3 would be very limited. The resulting likelihood function would have very small curvature in certain directions in the parameter space. Hence, although the resulting model would be locally identified almost everywhere, the parametrization would be inconvenient from the statistical inference viewpoint. Though a formal investigation of the issue is left for further research, there is an informal motivation for assumptions A1-A3. That is, crucial gains in terms of explanatory power are likely to stem from generalization of $p_\mathrm{u}(.)$ and $p_\mathrm{v}(.)$ rather than from relaxing A1-A3.

Once the parametric form of $p_u(.)$ and $p_v(.)$ is set, the vector of all statistical parameters is $\boldsymbol{\theta}' = \left(\boldsymbol{\beta} \ \boldsymbol{\theta}^{(v)\prime} \ \boldsymbol{\theta}^{(u)\prime}\right)'$. Statistical inference regarding $\boldsymbol{\theta}$ relies upon properties of the compound error term $\varepsilon_t$ which, for a production technology-type analysis ($\omega = 1$), is $\varepsilon_t = v_t - u_t$. In the general case its density $p_\varepsilon(.)$ is defined by the convolution of the respective densities of $v_t$ and $u_t$:

$$p_\varepsilon(\ln y - \ln g(\boldsymbol{x}; \boldsymbol{\beta}); \boldsymbol{\theta}^{(\varepsilon)}) = \int_{\mathbb{R}_+} p_v(\ln y - \ln g(\boldsymbol{x}; \boldsymbol{\beta}) + \omega u; \boldsymbol{\theta}^{(v)}) p_u(u; \boldsymbol{\theta}^{(u)}) du \qquad (2)$$

Note that in majority of practical applications $v_t$ is assumed to be Gaussian, while $p_u(.)$ is half-normal or exponential. Moreover, for now we assume that there exists one-to-one relationship between parameters of the structural form, i.e.: $\boldsymbol{\beta}, \boldsymbol{\theta}^{(v)}, \boldsymbol{\theta}^{(u)}$, and the parameters of the reduced form: $\boldsymbol{\beta}, \boldsymbol{\theta}^{(\varepsilon)}$, so no identification issues arise (possible identification problems are discussed later). However, in the case of stochastic frontier models, the statistical inference is not restricted to $\boldsymbol{\theta}$'s, since the latent variables ($u_t$'s), are also within the scope of interest, as they represent object-specific inefficiency terms.

We assume a general parametric distributional form for $p_v(.)$, and $p_u(.)$, making use of two distributions: the generalized $t$ distribution (GT; see MacDonald and Newey, 1988) and the generalized beta distribution of the second kind (GB2, see references in Harvey and Lange, 2017), respectively:

$$f_{GT}(v; \sigma_v, \nu_v, \psi_v) = \frac{1}{\sigma_v} \frac{\psi_v}{B(1/\psi_v, \nu_v/\psi_v) 2\nu_v^{1/\psi_v}} \left[\frac{1}{\nu_v}\left(\frac{|v|}{\sigma_v}\right)^{\psi_v} + 1\right]^{-(1+\nu_v)/\psi_v}, \qquad (3)$$

$$f_{GB2}(u; \sigma_u, \nu_u, \psi_u, \tau) = \frac{1}{\sigma_u} \frac{\psi_u}{B(\tau/\psi_u, \nu_u/\psi_u) \nu_u^{\tau/\psi_u}} \left(\frac{u}{\sigma_u}\right)^{\tau-1} \left[\frac{1}{\nu_u}\left(\frac{u}{\sigma_u}\right)^{\psi_u} + 1\right]^{-(\tau+\nu_u)/\psi_u}, \qquad (4)$$

with $B(.,.)$ denoting the beta function. The above formulation of the GT distribution assumes that the mode, median and mean (if the latter exists) is zero, in line with the usual formulation of SF-type models. The GB2 density is often formulated as $f_{GB2}(x; a, b, p, q) = \frac{|a| x^{ap-1}}{B(p,q) b^{ap}} \left[\left(\frac{x}{b}\right)^a + 1\right]^{-(p+q)}$. The parametrization in (4) used throughout this paper is its equivalent and implies: $p = \tau/\psi, q = \nu/\psi, a = \psi, b = \sigma \nu^{1/\psi}$. The reason for using an alternative parametrization is to emphasize the relationship between GT and GB2 distributions. Note that with $\tau = 1$, the $f_{GB2}(.; \sigma, \nu, \psi, 1)$ distribution is equivalent to half-GT distribution with parameters $f_{GT}(.; \sigma, \nu, \psi)$. In other words, the absolute value of a GT variable (distributed as $f_{GT}(.; \sigma, \nu, \psi)$) follows the GB2 distribution $f_{GB2}(.; \sigma, \nu, \psi, 1)$. However, the reason for not using the double GB2 (instead of GT) for $p_v(.)$ is that we would violate A1 (by double GB2 we mean an equal-weight mixture of $X_1$ and $-X_2$, with $X_i$ being independent copies of a GB2 variable with the same parameters). The double GB2 distribution with $\tau > 1$ would be bimodal, whereas $\tau < 1$ implies lack of continuity at the mode (with the density function approaching infinity from both sides of zero). We find such properties to be undesirable given the interpretation of the symmetric observation error term $v_t$ that is broadly accepted in SFA. However, this assumption could be relaxed within the framework described here as the resulting density of observables, $p_\varepsilon(.)$, would be nevertheless continuous.

It is clear that $\sigma_v$ in (3) and $\sigma_u$ in (4) are scale parameters (with $\sigma_v$ being analogous to the inverse of precision in Student's *t* distribution), while $\psi$'s and $\nu$'s are shape parameters. In particular, $\nu_u$ and $\nu_v$ control tail thickness (and, hence, the existence of finite moments – analogously to the *degrees-of-freedom* parameter in Student's *t* distribution). Moreover, for parameters $\nu_u$ and $\nu_v$ we also consider the limiting cases (of $\nu_u \to \infty$ or $\nu_v \to \infty$). In order to analyze the limiting behavior of the normalizing constants and the kernels of (3) and (4), note that (following Harvey and Lange, 2017; who cite Davis, 1964):

$$\frac{\psi}{B(\tau/\psi, \nu/\psi)\nu^{\tau/\psi}} \xrightarrow[\nu \to \infty]{} \frac{\psi}{\Gamma(\tau/\psi)\psi^{\tau/\psi}}, \tag{5}$$

$$\left[\frac{1}{\nu}\left(\frac{|x|}{\sigma}\right)^\psi + 1\right]^{-(\tau+\nu)/\psi} \xrightarrow[\nu \to \infty]{} \exp\left[-\frac{1}{\psi}\left(\frac{|x|}{\sigma}\right)^\psi\right], \tag{6}$$

where $\Gamma(.)$ denotes the gamma function. The above formulations indicate the relationship with the exponential family of densities, nested as limiting cases in the general model used here. In principle, it is also possible to consider the limiting behavior for $\psi_u$ or $\psi_v$, though these cases are less interesting from the empirical point of view given the usual interpretation in SFA, so this option is not considered here (e.g., $\psi_v \to \infty$ implies convergence towards uniform distribution on an interval). We would like to emphasize the importance of the following special cases:

1. As $\psi_v = 2$, the GT distribution (3) reduces to the Student's *t* distribution:

$$f_{ST}(z; \sigma, \nu) = \frac{1}{\sigma}\frac{2}{B(1/2, \nu/2)\nu^{1/2}}\left[\frac{1}{\nu}\left(\frac{z}{\sigma}\right)^2 + 1\right]^{-(1+\nu)/2}, \tag{7}$$

with $\sigma$ equivalent to $\sigma_v$ and $\psi$ equivalent to $\psi_v$ in (3); consequently, the GB2 distribution in (4) with $\tau=1$ and $\psi_u = 2$ reduces to the half-Student's *t* case.

2. As $\nu_v \to \infty$, the GT distribution (3) reduces to the Generalized Error Distribution (GED):

$$f_{GED}(z; \sigma, \psi) = \frac{1}{\sigma} \frac{\psi}{2\Gamma(1/\psi)\psi^{1/\psi}} \exp\left[-\frac{1}{\psi}\left(\frac{|z|}{\sigma}\right)^{\psi}\right], \tag{8}$$

consequently, the GB2 distribution with $\tau=1$, $\nu_u \to \infty$ reduces to half-GED.

3. The GB2 distribution (4) reduces to the generalized gamma distribution (GG) as $\nu_u \to \infty$:

$$f_{GG}(z; \sigma, \tau, \psi) = \frac{1}{\sigma} \frac{\psi}{\Gamma(\tau/\psi)\psi^{\tau/\psi}} \left(\frac{z}{\sigma}\right)^{\tau-1} \exp\left[-\frac{1}{\psi}\left(\frac{z}{\sigma}\right)^{\psi}\right]. \tag{9}$$

The original parametrization of the generalized gamma distribution according to Stacy (1962, eq. 1) is $f_{GG}(z; a, d, p) = \frac{p}{\Gamma(d/p)a^d} z^{d-1} \exp\left[-\left(\frac{z}{a}\right)^p\right]$. The relationship between the original parametrization and the one used in (9) is as follows: $d = \tau$, $p = \psi$, $a = \sigma\psi^{1/\psi}$.

Note that the Gaussian case can be obtained as a conjunction of the Student's *t* or GED sub-cases since it requires $\nu_v \to \infty$, and $\psi_v = 2$ (analogously, GB2 becomes half-normal with $\tau=1$, $\nu_u \to \infty$, and $\psi_u = 2$). Moreover, setting $\psi_v = 1$ instead of $\psi_v = 2$ leads to the Laplace distribution or the exponential case for GB2 with $\psi_u = \tau = 1$, and $\nu_u \to \infty$. The general SF model considered here is labeled GT-GB2, while its special case, which assumes $\tau=1$, is labeled GT-HGT (as the GB2 distribution is reduced into the half-GT case). Other special cases, with literature references, are listed in Table 1. Also, the density of $\varepsilon_t$ implied by (2), (3), and (4) in GT-GB2 is in general asymmetric and includes skew-t and skew-normal distributions as special cases.

[Table 1 about here]

## 4 Inference: likelihood-based methods for the GT-GB2 model

Parametric inference in the abovementioned models relies upon the likelihood function implied by (2)-(4) with latent variables integrated out (often referred to as integrated likelihood). Under our baseline assumptions (A1-A3), the observations are *i.i.d.* variables, so full log-likelihood for all the data can be trivially decomposed into a sum of observation-specific log-likelihoods of the general form:

$$\ln \mathcal{L}_t = A + C_t \tag{10}$$

$$A = \ln\left(\frac{1}{\sigma_v}\frac{1}{\sigma_u}\frac{\psi_v}{B\left(\frac{1}{\psi_v}, \frac{\nu_v}{\psi_v}\right)2\nu_v^{\frac{1}{\psi_v}}}\frac{\psi_u}{B\left(\frac{\tau}{\psi_u}, \frac{\nu_u}{\psi_u}\right)\nu_u^{\frac{\tau}{\psi_u}}}\right) \tag{11}$$

$$C_t = \ln\left(\int_0^{+\infty} \left[\frac{1}{\nu_v}\left(\frac{|\ln y_t - \ln g(x_t; \beta) + u|}{\sigma_v}\right)^{\psi_v} + 1\right]^{-\frac{(1+\nu_v)}{\psi_v}} \left(\frac{u}{\sigma_u}\right)^{\tau-1} \left[\frac{1}{\nu_u}\left(\frac{u}{\sigma_u}\right)^{\psi_u} + 1\right]^{-\frac{(\tau+\nu_u)}{\psi_u}} du\right) \quad (12)$$

This is not given in a closed form, as the integral in (12) cannot be solved analytically in the general case. However, as the integral is one-dimensional, it can be evaluated with arbitrary precision for any given data point and parameter vector using numerical methods (see, e.g., Shampine, 2008). Hence, with $T$ observations, one evaluation of the log-likelihood requires $T$ computations of the above one-dimensional integral. The problem can be parallelized, so given modern techniques the computational cost is not prohibitive even in the case of simulation-based estimation techniques relying on many evaluations of the likelihood. Obviously, in the case of Bayesian estimation, the integrated likelihood should be multiplied by priors for model parameters. Further numerical aspects of the inference are discussed in Appendix 1.

Due to the direct use of integrated likelihood (10)-(12), point estimation requires the usual optimization over the parameter space (maximization of the log-likelihood or log-posterior). The results allow for computation of Bayesian Information Criterion (BICs), or BIC-based approximate Bayes Factors and the corresponding model weights (though one might prefer to use an approximate LR-like model reduction testing procedure instead). If an estimate of covariance matrix (e.g. via numerical Hessian of log-posterior) is available, it is possible to compute Bayes Factors via Laplace approximation, which is simple and reliable. Hence, crucially, it is possible to conduct some form of formal model comparison without the use of computationally intensive procedures, which is rather unusual when dealing with non-Gaussian models with latent structures.

As for interval estimation, one option is the use of approximations based on multivariate normality in the parameter space. Properties of Maximum Likelihood (ML hereafter) estimation of models of similar degree of complexity - though allowing for closed-form likelihoods - are discussed by Harvey and Lange (2017). However, our empirical results indicate that indeed the multivariate normal approximation is not necessarily relevant in the case of shape parameters ($\psi_u, \psi_v, \nu_u, \nu_v, \tau$). Therefore, in this paper we make use of more elaborate Bayesian inference techniques to deal with the issue. For this purpose we use informative priors (to ensure the existence of posterior) which, informally, can be considered as close to uninformative (details are discussed in Appendix 1). However, the goal could also be obtained using, e.g., bootstrap methods.

Bayesian models are quite widespread in SFA because Bayesian paradigm provides convenient solutions to the problem of statistical inference on latent variables (i.e., inefficiencies). However, standard numerical tools for Bayesian inference used in well-known SF models display certain drawbacks. In particular, the most widespread methods of Bayesian estimation of SF models are based on Gibbs sampling (see Osiewalski and Steel, 1998; Griffin and Steel, 2007). Such sampler is very efficient (in terms of computational power) and makes it possible to simulate the statistical parameters jointly with the latent variables in a single Markov chain (MCMC). However, its successful application usually relies upon certain simplifying assumptions as to the form of priors (being quasi-conjugate to form full conditional densities of standard form) as well as the frontier. Since our goal is to develop a framework that is as general as possible (within the parametric approach), we follow an alternative path, using general-purpose MCMC algorithms instead of model-specific Gibbs sampling schemes. This strategy does not require linearity of the frontier with respect to parameters or priors belonging to certain narrow parametric classes. Estimation of statistical parameters of the GT-GB2 model can be carried out by a more general tool such as Metropolis-Hastings algorithm, which does not necessitate any specific form of likelihood or prior. In such case, $\theta's$ can be sampled in a single MCMC chain (without u's forming a T-dimensional latent variable). This approach might be less efficient in terms of computational power but it is also far less restrictive. Hence, the estimation strategy advocated here, though computationally demanding, has three important advantages over Gibbs sampling-based estimation in SFA:

- it does not require specific, "convenient" forms of $p_v(.)$ or $p_u(.)$ - Gibbs samplers have been developed only for some of special cases listed in Table 1;
- it does not require specific classes of priors, so it is possible to conduct an in-depth analysis of prior sensitivity and prior coherence;
- it does not rely on any specific (e.g. linear or loglinear) form of the frontier $g(x_t; \beta)$; although our empirical applications in Section 6 follow previous studies and make use of the traditional loglinear forms of $g$, this aspect is worth noting as it further adds to the generality of the proposed framework.

As u's are not drawn within the same Markov chain, the resulting sampler is likely to display better mixing properties. Moreover, based on estimation results for $\boldsymbol{\theta}'s$, it is possible to draw u's in a separate MCMC run. Details regarding inference on u's are provided in Appendix 1.

For many applications, the general GT-GB2 specification described here is likely to be overparametrized. Thus, in our view the model may serve as a platform for specification search as it is not obvious which reduced cases are relevant for the data at hand. Consequently, it is likely that in practice it would be preferable to use approximate inference procedures. Indeed, it is possible to run a sequence of maximum-posterior (or ML) estimates over the number of nested special cases listed in Table 1 and the computational cost of such procedure is relatively small. Such a sequence of point estimates can be used either for model comparison or model selection (based on quasi-LR tests, BIC values or BIC/Laplace approximation-based Bayes factors and model posterior probabilities). If the results indicate one clearly superior model, full MCMC inference can be conducted for this case only. Alternatively, it is possible to run MCMC over a couple of models with non-negligible posterior probabilities and combine the results using Bayesian model averaging (BMA). Such approximate strategies might be sufficient to deal with most model uncertainty problems. Obviously, it is possible to perform full BMA over the whole range of models. However, approximate procedures, like the one mentioned above, may be more practical.

Moreover, if a researcher is making use of an existing SF model, our approach provides a relatively simple check to verify whether, for the dataset at hand, the specification used is empirically valid (via examining e.g. LR-type ratio versus the GT-GB2 model). We therefore emphasize that the key advantage of our approach is that it allows for fairly simple model comparison over a very broad and flexible model class. The increasing computational power – with an emphasis on parallelization-friendly problems, as the one considered here – makes the proposed framework a very prospective approach to SFA. This is because while the computational burden is likely to decrease in time, the benefits will remain.

# 5 Possible model extensions

## 5.1 Varying Efficiency Distribution

The basic SFA formulation, labeled Common Efficiency Distribution (CED-SFA), does not allow for heterogeneity (across observations) in the inefficiency process. This assumption is relaxed within the Varying Efficiency Distribution model class (VED-SFA; see, e.g., Koop et al., 1997). VED-SFA models allow exogenous variables to influence the inefficiency process. They are characterized by the fact that the inefficiency distribution depends upon certain covariates (hence, u's are independent but no longer identically distributed); see Koop et al. (1997). Within the GT-GB2 framework a natural option is to replace $\sigma_u$ in (4) with:

$$\sigma_{u,t} = \exp(\gamma + \boldsymbol{\delta w}_t), \qquad (13)$$

where $\boldsymbol{w}_t$ represent a vector of covariates driving differences across objects with respect to inefficiency distribution, $\gamma$ and $\boldsymbol{\delta}$ are model parameters replacing $\sigma_u$. Importantly, such a general formulation carries over to VED-type versions of all nested special cases listed in Table 1 – which results in a whole new class of coherent VED-SFA formulations.

In principle, it is possible to extend the idea to consider individual effects in shape parameters as well. However, it is not obvious whether such formulation would turn out to be empirically relevant. From empirical viewpoint, an important distinction would be between covariate-driven $\sigma_{u,t}$ (individual effects in inefficiency terms) and covariate-driven $\sigma_{v,t}$ (heteroskedasticity in observation error). Such formulations are fully feasible extensions to the GT-GB2 framework presented here.

## 5.2 Panel data modelling

A number of proposals have been made to account for panel data structure in SFA. Here we comment on the most common cases, which can be relatively easily implemented in the GT-GB2 framework presented in the paper.

First, we can consider u's as time-invariant effects (or object-specific; see, e.g., Model I in Pitt and Lee, 1981). This is the most traditional setting in which inefficiency is to capture persistent effects that differentiate objects' performances. In other words, all time-invariant differences in performances are attributed to inefficiency and all transient effects are treated as part of random disturbance ($v_{it}$). The corresponding production-oriented SFA model ($\omega = 1$) is

$$\ln y_{it} = \ln g(x_{it}; \beta) + v_{it} - u_i \tag{14}$$

where $i$ indices an object ($i = 1, \ldots, n$) and $t$ is the time index ($t = 1, \ldots, T$). The resulting general convolution of densities for $v_{it}$ and $u_i$ is then

$$\begin{aligned} f(\varepsilon) &= \prod_{i=1}^{n} \prod_{t=1}^{T} p_\varepsilon \left( \ln y_{it} - \ln g(x_{it}; \beta); \theta^{(\varepsilon_{it})} \right) \\ &= \prod_{i=1}^{n} \left\{ \int_{R+} \left\{ \prod_{t=1}^{T} p_v (\ln y_{it} - \ln g(x_{it}; \beta) + u_i) \right\} p_u \left( u_i; \theta^{(u_i)} \right) du \right\}. \end{aligned} \tag{15}$$

This results in a likelihood function similar to that in (10). The change is that now we are down to evaluating only $n$ integrals. On the other hand, however, each integral needs to be calculated over a product of $T$ densities $p_v$. This is more challenging in terms of fine-tuning the integration procedure (specifying relevant waypoints etc.) and becomes increasing complex as $T$ increases. We reckon, however, that this is still not as time consuming as having to evaluate $nT$ integrals in the baseline (pooled) model. So the net effect is likely an increase in computational speed.

Another popular way is to add another latent variable ($\alpha_i$) to represent an object-specific effect much like in the True Random Effects SF model (Greene, 2004). In this setting inefficiency is to capture transient effects that differentiate objects' performances over time, while the random effect $\alpha_i$ takes on the persistent part. Thus, time-invariant differences are attributed to heterogeneity of the frontier/technology:

$$\ln y_{it} = \ln g(\boldsymbol{x_{it}}; \boldsymbol{\beta}) + v_{it} - u_{it} + \alpha_i. \tag{16}$$

Assuming that $\alpha_i \sim f_N(\alpha_i; 0, \sigma_\alpha^2)$ and that $\sigma_\alpha^2$ has the usual inverse-gamma prior $\sigma_\alpha^{-2} \sim f_G(\sigma_\alpha^{-2}; 0.5 \cdot 10^{-4}, 0.5 \cdot 10^{-4})$ we can use Bayesian inference to estimate the model. In particular we can employ the following sampling scheme; set initial value for vector $\alpha$ (e.g., zero) and then for i=1 to S:

1) draw parameters using Eq. (11) and Eq. (16) conditional on $\alpha$ taken from step (4) in draw i-1;
2) draw u's according to Eq. (A1.1) in Appendix 1 using values in step (1) from the current draw and step (4) taken from draw i-1;

3) draw $\sigma_\alpha$ from the conditional $p(\sigma_\alpha^{-2}|y, X, \theta_{-\sigma_\alpha}) = f_G(\sigma_\alpha^{-2}|0.5(10^{-4} + n), 0.5(10^{-4} + \alpha'\alpha))$ using $\alpha$ from step (4) in draw i-1;

4) draw $\alpha$'s from the conditional $p(\alpha|data, \theta_{-\alpha}) = f_N^n(\alpha | \frac{\sigma_\alpha^2}{\frac{\sigma_v^2}{T} + \sigma_\alpha^2} \tilde{\alpha}, \frac{\frac{\sigma_v^2 \sigma_\alpha^2}{T}}{\frac{\sigma_v^2}{T} + \sigma_\alpha^2} I_n)$ based on (1-3); where $\tilde{\alpha} = \ln g(\overline{X}; \beta) - \ln \overline{y} + \overline{u}$ and symbol " $\overline{\phantom{x}}$ " denotes an $n$-element vector of $n$ object-wise averages (averages over time for $n$ objects) for $\ln y$, $X$, and u.

Similarly to traditional Gibbs sampling, after an initial burn-in phase the procedure samples from the sought-after posterior distribution and we can use the accepted draws to approximate any model quantity of interest. This procedure is more time consuming as it requires us to draw $u$ (note that in the baseline specification we don't need to estimate $u$ in order to evaluate the model and, e.g., perform model search). However, this is the price to pay for a more complex model structure.

A third option - probably the easiest to implement - is to follow approach based on True Fixed Effects SF model (Greene, 2004). From a Bayesian perspective modelling fixed effects amounts to having each prior on $\alpha_i$ with a different scale parameter (e.g., $\sigma_{\alpha_i}$). Assuming that $\alpha_i \sim N(0, \sigma_{\alpha_i}^2)$ we can simplify the problem by considering the intercept, e.g. $\beta_0 \sim N(0, \sigma_{\beta_0}^2)$, and the fixed effect jointly as $\delta_i = \beta_0 + \alpha_i$. Hence, in practice this approach amounts to estimating an object-specific intercept $\delta_i$ of the frontier under the following prior: $\delta_i \sim N(0, \sigma_{\alpha_i}^2 + \sigma_{\beta_0}^2 + 2\rho_i \sigma_{\alpha_i} \sigma_{\beta_0})$, where $\rho_i$ it the correlation between $\alpha_i$ and $\beta_0$. Since we usually assume independence between $\alpha_i$ and $\beta_0$ the prior on $\delta_i$ simplifies to $N(0, \sigma_{\alpha_i}^2 + \sigma_{\beta_0}^2)$.

There are many other panel data modelling techniques proposed within the SFA literature (see, e.g., Parmeter and Kumbhakar, 2014; for an overview). However, the abovementioned ones can be relatively easily incorporated within the proposed modelling framework.

## 6 Empirical applications

Empirical applications are based on the following well-researched datasets, extensively covered in many papers in the field of (in)efficiency analysis:

(i) World Health Report by WHO (see, e.g., WHO, 2000; Tandon et al., 2000; Evans et al., 2000; Hollingsworth and Wildman, 2002; Greene, 2004; 2005; 2017);

(ii) Spanish dairy farms (see, e.g., Cuesta, 2000; Alvarez, et al. 2006, Greene, 2017).

For space considerations more detailed results are provided for the first dataset (WHO), while for the latter we report only basic results for model specification search and averaging (detailed results are available in Appendix 2). Furthermore, we note that BMA-type model averaging is usually applied in order to ensure adequate choice of exogenous variables (within the same stochastic structure; see, e.g., Makieła and Osiewalski, 2018). Our use of BMA here is different; i.e., we fix the set of exogenous variables and average results over competing stochastic specifications as this reflects actual problems of model uncertainty in SFA.

In this section we also make an additional – empirically driven – assumption: (A4) $u_t$'s have non-increasing pdf's. We note that there is a considerable SFA literature about non-monotonic distributions of $u_t$'s, which would indicate their usefulness (see, e.g., Stevenson, 1980; Greene, 1990; van den Broeck et al., 1994; Griffin and Steel, 2004, 2007, 2008; among many). Such applications, however, have assumed restrictive distributional forms of $v_t$'s, which could be the reason why non-monotonic distributions of $u_t$'s have been found relevant in the first place (e.g., due to outliers; see Stead et al., 2018; for a discussion about outliers detection and treatment). In our view, generalization of distributional assumptions as to $v_t$ in empirical applications is sufficient to maintain similar statistical fit despite the use of strictly non-increasing distribution of $u_t$. Also, having both generalized distributions of $v_t$ and non-monotonic distributions of $u_t$ in the same model, may yield statistical identification problems which already motivated assumptions A1-A3. Nevertheless, we emphasize that A4 is an additional assumption (or model restriction) used in this section, not a model requirement itself.

### 6.1 World Health Report

The dataset used in this section is based on WHO (2000) study which contains annual information on healthcare attainments in 191 countries in 1993-1997. The dataset has been extensively described in the aforementioned papers. We use countries with complete five year observations, which is 140 countries and 700 observations in total. Given the above-listed studies we make use of the following translog function:

$$y_{it} = \beta_0 + \beta_1 x_{ti,1} + \beta_2 x_{ti,2} + \beta_3 x_{ti,1}^2 + \beta_4 x_{ti,2}^2 + \beta_5 x_{it,1} x_{it,2} + v_{it} - u_{it} \qquad (17)$$

where y is the log of COMP (a composite measures of success in achieving hive health goals: (1) health, (2) health distribution, (3) responsiveness, (4) responsiveness in distribution and (5) fairness in financing); $x_1$ is the log of health expenditure per capita, $x_2$ is the log of education attainment (average years of schooling), while $i=1,...,140$ and $t=1,...,5$ are country and year indices respectively. Some previous studies would drop parts of the full transolg model; e.g., Tandon et al. (2000) and Evans et al. (2000) would have $\beta_3 = \beta_5 = 0$, while Greene (2005) would use a Cobb-Douglas form ($\beta_3 = \beta_4 = \beta_5 = 0$). Evidently, research on the form of the frontier is also of much interest in SFA and it could be used within this framework. In particular, we can consider complex nonlinear forms of the frontier in BMA because our inference method does not require any specific form of $g(.)$; see Section 4. Nonetheless, as mentioned earlier we leave the frontier parametrization fixed in order to focus our investigation on SF model specification uncertainty (with respect to the stochastic components v and u).

Table 2 shows results of model comparison based on GT-GB2 and other 33 distinctive SF models, all of which are special cases of GT-GB2. We compare the approximate results based on ML (column label $w_3$) and the popular Bayesian Information Criterion (BIC) with exact Bayesian results under two prior model probabilities: (i) a uniform prior which gives each model equal prior probability (1/34) to every model (column label $w_1$) and (ii) a prior that follows the rule of Ockham Razor and prefers parsimonious models (column label $w_2$). That is, we take $p(M_i) \propto 2^{-k}$ where $k$ is the number of parameters. Hence, a model with one additional parameter is *a priori* two times less likely than its more parsimonious counterpart. This particular prior was first used by Osiewalski and Steel (1993) so we refer to it as the "OS prior" hereafter.

Based on the results in Table 2 we find that approximate ML results overinflate the significance of the best model by assigning it model weights roughly about 0.957-0.979 as compared to 0.426-0.640 based on Bayesian inference. Model rankings between the two methods are largely similar especially if we consider the Bayesian results with OS prior (correlation is about 0.964).

[Table 2 about here]

The best model specification is the one in which inefficiency follows distribution similar to generalized gamma (GG), GB2 being its probable extension given Bayesian results. The observation error is likely Laplace-shaped. Thus, LAP-GG is the favored specification given the data. Even though Bayesian results also point towards five other nonnegligible specifications we should note that these are also very similar to LAP-GG. Furthermore, regardless of the method used (ML or Bayes) models with half-normal distribution of inefficiency are by far the worst specifications. E.g., the most popular normal-half-normal SF model is roughly nine orders of magnitude ($10^9$) less likely given the data than its somewhat less popular normal-exponential counterpart. Similar differences can be found in other models employing half-normal distribution of inefficiency.

Table 3 presents model averaging results for the stochastic parameters based on full Bayesian inference (results for the frontier parameters are provided in Appendix 2). We can notice that estimates of the pooled model only slightly deviate from the abovementioned LAP-GG specification. Also, given results in Table 4 we notice that $P(\tau = 1|y)$, $P(\psi_u = 1|y)$ and $P(\psi_u = 2|y)$ are virtually zero while $P(v_u \to \infty|y)$ is about 0.6-0.77. This indicates that GG-type distribution of inefficiency is clearly preferred by the data though the more general GB2 distribution is not without merit. The estimates of scale parameters are relatively consistent across all models and there seems to be no relevant posterior relation between $\sigma_u$, $\sigma_v$ and $\tau$ (see Figure 1).

[Table 3 and 4 about here]

[Figure 1 about here]

Table 5 provides a summary of efficiency ($r = \exp(-u)$) comparison across all models (correlation matrix can be viewed in Appendix 2). In general correlations between efficiency scores across all models are high. So, if we were only to rank observations based on efficiency scores then surprisingly the simplest – and least adequate – models would probably suffice. However, we note substantial differences in terms of average levels and ranges in terms of efficiency between specifications (i.e., the differences between the most and least efficient observations). Since exact values of efficiencies are often directly linked with performance-oriented policies such differences can translate into considerable losses or gains for individual agents. Furthermore, the best models are the ones which indicate relatively high average efficiencies with relatively narrow range (i.e., differences between the most and least efficient unit). This is in line with previous findings in this regard (Makieła and Osiewalski, 2018).

[Table 5 about here]

Finally, Figure 2 shows posterior density plot of efficiency in the pooled model under both priors. We notice that in case of WHO data the shape remains unchanged regardless of the prior model probability used. Also, the distribution is clearly bi-modal with one of the modes being at one (indicating full efficiency). This may provide some evidence in favor of a zero-inefficiency stochastic frontier model specifications (see, e.g., Kumbhakar et al., 2013).

[Figure 2 about here]

## 6.2 Spanish dairy farms

The dataset contains information on 247 farms in 1993-1998. Again, following the literature we use translog production function:

$$y_{it} = \beta_0 + \beta_1 t + \sum_{j=2}^{J+1} \beta_j x_{ti,j} + \sum_{j}^{J+1} \sum_{g \geq j}^{J+1} \beta_{j,g} x_{it,j} x_{it,g} + v_{it} - u_{it} \qquad (18)$$

where y is the log of milk output, x is a vector of logs of inputs (number of milk cows, land, labor and feed), *J* denotes the number of inputs (*J* = 4) while *i*=1,...,247 and *t*=1,…,6 are object and time indices respectively. Thus we have 16 elements in vector *β* (15 slope parameters and the intercept) and 1482 observations.

Table 6 summarizes results of the model specification search (estimates of model parameters are available in Appendix 2). We can notice that given the exact Bayesian results there are two relevant model specifications: GT-GB2 and Laplace-half-normal (LAP-HN). The two models have roughly equal explanatory power as measured by the marginal data density ($\ln p(y)$: 810.518 vs. 810.495) so their posterior probabilities largely depend on the prior. Under equal prior odds (uniform prior) the two models are equally relevant; but under OS prior the simpler model clearly dominates the ranking. This effect is further strengthened in case of approximate ML results. Again, we find high correlation of model probability ranking between ML and Bayes (especially with OS prior: 0.956 correlation) and that approximate ML results assign much higher model weights to the simpler LAP-HN model. Evidently ML over-penalizes larger models.

[Table 6 about here]

Similarly to WHO example the data prefer a Laplace-shaped observation error. However, the inefficiency term this time is likely to be close to half-normality, with an exponential case being significantly less probable (see Table 7). A more detailed investigation regarding this can be made using the full GT-GB2 model. Figure 3 presents heatmaps of 2D marginal distributions for $\sigma_u$, $\sigma_v$ and $\tau$. We notice two 'regimes' for scale parameters: one for $\tau$ around 0.2 (implying generalized gamma-type distributions of inefficiency like in WHO case); and one for $\tau = 1$ (implying half-GT/GED/normal class of distributions). Given the results in Table 7 the notion that $\tau = 1$ cannot be rejected. We do note, however, a distinctive difference between the two model classes with respect to scale parameters.

[Table 7 and Figure 3 about here]

Compared to WHO example the model ranking is much more even. However, this is somewhat to be expected if we look at the differences between the highest and the lowest recorded maximum log-likelihood values for the two datasets. In case of WHO it is about 49.2 while for Spanish dairy farms it is 'only' 14.8. Similar evidence can be found when comparing the log of marginal data densities ($\ln p(y)$). Hence, no surprise that the posterior model ranking is more even in case of Spanish dairy farms and the impact of the prior model probability much more substantial. As a result, the posterior distribution of efficiency in the pooled model is driven by the prior model probability (see Figure 4).

[Figure 4 about here]

# 7 Conclusions

The main contribution of this paper comes with the development of a new stochastic frontier model, labeled GT-GB2, which generalizes almost all parametric specifications used in the field. It is based on flexible assumptions regarding the compound error, with sampling density defined as a convolution of two densities: generalized *t* (for observation error) and generalized beta of the second kind (for inefficiency). Consequently, it allows for a broad range of deviations from the popular normal-half-normal or normal-exponential models, while maintaining some crucial regularity conditions described in the paper. Due to its parametric structure the model provides a natural generalization towards the VED-like family and can be further redeveloped for panel data treatment, as discussed in Section 5. It is also possible to generalize our approach along the lines of ZISF-type models (see Kumbhakar et al. 2013), for example using the results of Harvey and Ito (2020), which is subject to further research.

We also develop exact and approximate inference methods (using integrated likelihood) which allow for model averaging. This is important for addressing model specification uncertainty with respect to inference on latent variables. These variables (or their functions) represent individual inefficiency ($u$) or efficiency ($r$) terms, which are key quantities of interest in SFA. In order to provide adequate description of all aspects of their estimation uncertainty, we use full Bayesian inference with BMA, which is rather costly in terms of computational power. However, our model framework also allows for approximate – yet prior-free – approach to density estimation of latent variables and model averaging. By avoiding the use of simulation-intensive methods this can be a convenient alternative to full BMA.

The Bayesian inference strategy is well feasible given contemporary computation power of a PC. We have shown that with $T = 1482$ observations and a heavily-parametrized frontier (16 parameters in vector $\beta$), it is possible to estimate the general GT-GB2 model with its nested sub-cases and to conduct model averaging procedures. Note that we are able to overcome two of key limitations of the Bayesian approach that are often encountered in practice. First, we develop a simple yet reliable method for computation of marginal data densities (hence: posterior model probabilities which are essential for BMA) via Laplace approximation. In non-Gaussian models with latent variables this solution is usually infeasible while reliable alternatives are difficult to obtain. Second, contrary to the popular Gibbs sampling our approach can be used with any choice of (proper) priors, so it is possible to develop coherent priors or to conduct an in-depth analysis of prior sensitivity (see, e.g., Makieła and Mazur, 2020). The latter is important in order to convince other researchers that the results are not prior-driven. Since this is a very specific and purely Bayesian problem, likely of limited interest to SFA practitioners, we leave the derivation of optimal (or reference) priors within the GT-GB2 class for further considerations.

In the approximate approach point estimates of parameters can be obtained by maximizing the log-likelihood or log-posterior using the available integrated likelihood. Crucially, the results are less computationally demanding and sufficient for model comparison (i.e., computation of model weights for model averaging). This amounts to the use of BICs or the aforementioned Laplace approximation (the latter requires slightly more effort when numerical Hessian of log-posterior is unavailable). We also suggest an approximate method for density estimation of latent variables (see Appendix 1). This approach avoids numerically intensive MCMC methods and allows to estimate inefficiency using model averaging at a lower computational cost.

Our empirical findings in Section 6 indicate that the symmetric disturbance is likely to follow a Laplace class of distributions (see, e.g., Horrace and Parmeter, 2018). It seems that it is important to take into account the Laplace distribution when testing for heavy tails in the symmetric term (so the alternatives of Gaussian vs. Student's *t* are too restrictive). However, there is no particular indication as regards the distribution of inefficiency. So far, inefficiency distribution seems to be data-specific and thus GB2 distribution is a convenient basis for inefficiency specification search.

Correlations of model probability rankings based on ML and Bayesian inference are quite similar especially under model prior probabilities that favor parsimony; though we must point out that ML model weights penalize larger models much more radically (as the model weights used here are BIC-based). Moreover, there is a growing tendency in SFA to use heavy tailed distributions (see, e.g., Horrace and Wang, 2020). Our findings suggest that in determining a given model's adequacy it is more important how its stochastic components' distributions behave around the mode than in the tails (or to generalize beyond the use of *t*-distribution for the symmetric term). Heavy tails can be easily tested within our framework along with other features of the components' distributions.

Since one of key advantages of the GT-GB2 SF model class is its flexibility, one might be tempted to consider its non-parametric alternatives. In particular, an interesting option is that of Florens et al. (2019). However, despite some obvious limitations of the parametric approach, it is relatively straightforward to impose assumptions A1-A3 as well as A4 within the GT-GB2 model; something which is non-trivial within the non-parametric framework. Moreover, the basic GT-GB2 model can be augmented in order to account for covariate-dependent scale/shape characteristics (along the lines suggested in Section 5), which again might be difficult to achieve within the non-parametric setup. Nonetheless, the parametric framework considered in this paper provides a research strategy that is complementary to that of Florens et al. (2019).

# Tables and Figures

Table 1. Selected models nested in GT-GB2 SF specification

| $p_v(v;.)$ | $p_u(u;.)$ | Restrictions in GT-GB2 SF | Reference | Model label |
|---|---|---|---|---|
| normal | half-normal | $\nu_u, \nu_v \to \infty, \psi_u = \psi_v = 2, \tau = 1$ | Aigner et al. (1977) | N-HN |
| normal | exponential | $\nu_u, \nu_v \to \infty, \psi_u = 1, \psi_v = 2, \tau = 1$ | Meeusen & Broeck (1977) | N-EXP |
| normal | gamma | $\nu_u, \nu_v \to \infty, \psi_u = 1, \psi_v = 2$ | Greene (1990) | N-GAM |
| normal | Weibull | $\nu_u, \nu_v \to \infty, \psi_v = 2; \psi_u = \tau$ | Tsionas (2007) Griffin & Steel (2008) | LAP-W |
| normal | gen.-gamma | $\nu_u, \nu_v \to \infty, \psi_v = 2$ | Griffin & Steel (2008) | N-GG |
| Student's $t$ | half-Student's $t$ | $\psi_u = \psi_v = 2, \tau = 1$ | Tancredi (2002)** Tchumtchuoa & Day (2007)** | T-HT |
| Student's $t$ | half-normal | $\nu_u \to \infty, \psi_u = \psi_v = 2, \tau = 1$ | Griffin & Steel (2007)* Stead et al. (2019) | T-HN |
| Student's $t$ | exponential | $\nu_u \to \infty, \psi_u = 1, \psi_v = 2, \tau = 1$ | Griffin & Steel (2007)* | T-EXP |
| Student's $t$ | gamma | $\nu_u \to \infty, \psi_u = 1, \psi_v = 2$ | Griffin & Steel (2007)* | T-GAM |
| Student's $t$ | GB2 | $\psi_v = 2$ | NEW | T-GB2 |
| Student's $t$ | half-generalized $t$ | $\psi_v = 2, \tau = 1$ | NEW | T-HGT |
| Student's $t$ | gen.-gamma | $\nu_u \to \infty, \psi_v = 2$ | NEW | T-GG |
| Student's $t$ | half-GED | $\nu_u \to \infty, \psi_v = 2, \tau = 1$ | NEW | T-HGED |
| $t$-Laplace | GB2 | $\nu_u \to \infty, \psi_v = 1$ | NEW | TLAP-GB2 |
| Laplace | exponential*** | $\nu_u, \nu_v \to \infty, \psi_u = \psi_v = 1, \tau = 1$ | Nguyen (2010) Horrace & Parmeter (2018) | LAP-EXP |
| Laplace | half-normal | $\nu_u, \nu_v \to \infty, \psi_u = 2, \psi_v = 1, \tau = 1$ | NEW | LAP-HN |
| Laplace | half-Student's $t$ | $\nu_v \to \infty, \psi_u = 2, \psi_v = 1, \tau = 1$ | NEW | LAP-HT |
| Laplace | half-GED | $\nu_u, \nu_v \to \infty, \psi_v = 1, \tau = 1$ | NEW | LAP-HGED |
| Laplace | Gamma | $\nu_u, \nu_v \to \infty, \psi_u = \psi_v = 1$ | NEW | LAP-GAM |
| Laplace | half-generalized $t$ | $\nu_v \to \infty, \psi_v = 1, \tau = 1$ | NEW | LAP-HGT |
| Laplace | gen.-gamma | $\nu_u, \nu_v \to \infty, \psi_v = 2$ | NEW | LAP-GG |
| GED | GB2 | $\nu_v \to \infty$ | NEW | GED-GB2 |
| GED | half-generalized $t$ | $\nu_v \to \infty, \tau = 1$ | NEW | GED-HGT |
| GED | half-Student's $t$ | $\nu_v \to \infty, \psi_u = 2, \tau = 1$ | NEW | GED-HT |
| GED | half-GED | $\nu_u, \nu_v \to \infty, \tau = 1$ | NEW | GED-HGED |
| GED | gen-gamma | $\nu_u, \nu_v \to \infty$ | NEW | GED-GG |
| GED | Gamma | $\nu_u, \nu_v \to \infty, \psi_u = 1$ | NEW | GED-GAM |
| GED | Weibull | $\nu_u, \nu_v \to \infty, \psi_u = \tau$ | NEW | GED-W |
| GED | half-normal | $\nu_u, \nu_v \to \infty, \psi_u = 2, \tau = 1$ | NEW | GED-HN |
| GED | Exponential | $\nu_u, \nu_v \to \infty, \psi_u = 1, \tau = 1$ | NEW | GED-EXP |
| Generalized $t$ | Exponential | $\nu_u \to \infty, \psi_u = 1, \tau = 1$ | NEW | GT-EXP |
| Generalized $t$ | half-normal | $\nu_u \to \infty, \psi_u = 2, \tau = 1$ | NEW | GT-HN |
| Generalized $t$ | half-Student's $t$ | $\psi_u = 2, \tau = 1$ | NEW | GT-HT |
| Generalized $t$ | half-GED | $\nu_u \to \infty, \tau = 1$ | NEW | GT-HGED |
| Generalized $t$ | half-generalized $t$ | $\tau = 1$ | NEW | GT-HGT |
| Generalized $t$ | gen.-gamma | $\nu_u \to \infty$ | NEW | GT-GG |

Note: * these models were only briefly discussed; ** only models with common degrees of freedom parameter (i.e. $\nu_u = \nu_v$) where considered; *** the paper discusses truncated Laplace but only exponential is used in estimation; all of the models listed (apart from n-w, ged-w and ged-gam) are included in the empirical study.

Table 2. WHO: model comparison results

| Model label | $w_1$ | $w_2$ | $w_3$ | $\ln p(y\|M_i)$ | $\ln p_{ML}(y\|M_i)$ | $\ln L$ | BIC | $k$ |
|---|---|---|---|---|---|---|---|---|
| GT-GB2 | 0.1320 | 0.0495 | 0.0000 | 922.156 | 919.718 | 962.300 | -1839.44 | 13 |
| LAP-GG | 0.4259 | 0.6395 | 0.9566 | 923.327 | 929.586 | 962.341 | -1859.17 | 10 |
| TLAP-GB2 | 0.2005 | 0.1506 | 0.0000 | 922.574 | 923.035 | 962.341 | -1846.07 | 12 |
| GT-GG | 0.0925 | 0.0695 | 0.0015 | 921.801 | 923.095 | 962.401 | -1846.19 | 12 |
| GED-GG | 0.0855 | 0.0642 | 0.0384 | 921.722 | 926.370 | 962.401 | -1852.74 | 11 |
| GED-GB2 | 0.0571 | 0.0214 | 0.0015 | 921.318 | 923.109 | 962.416 | -1846.22 | 12 |
| T-GB2 | 0.0061 | 0.0045 | 0.0000 | 919.075 | 920.745 | 960.051 | -1841.49 | 12 |
| N-GG | 0.0005 | 0.0007 | 0.0005 | 916.519 | 922.092 | 954.847 | -1844.18 | 10 |
| LAP-GAM | 0.0000 | 0.0000 | 0.0000 | 908.821 | 915.841 | 945.321 | -1831.68 | 9 |
| GT-EXP | 0.0000 | 0.0000 | 0.0000 | 903.092 | 908.027 | 940.782 | -1816.05 | 10 |
| GED-EXP | 0.0000 | 0.0000 | 0.0000 | 902.737 | 911.303 | 940.782 | -1822.61 | 9 |
| LAP-EXP | 0.0000 | 0.0000 | 0.0000 | 901.478 | 910.809 | 937.013 | -1821.62 | 8 |
| GT-HGT | 0.0000 | 0.0000 | 0.0000 | 901.301 | 904.158 | 943.465 | -1808.32 | 12 |
| GT-HGED | 0.0000 | 0.0000 | 0.0000 | 901.198 | 907.438 | 943.469 | -1814.88 | 11 |
| GED-HGED | 0.0000 | 0.0000 | 0.0000 | 900.858 | 910.709 | 943.464 | -1821.42 | 10 |
| GED-HGT | 0.0000 | 0.0000 | 0.0000 | 900.638 | 907.435 | 943.466 | -1814.87 | 11 |
| T-EXP | 0.0000 | 0.0000 | 0.0014 | 900.310 | 906.087 | 935.567 | -1812.17 | 9 |
| LAP-HGED | 0.0000 | 0.0000 | 0.0000 | 899.958 | 911.223 | 940.703 | -1822.45 | 9 |
| N-EX | 0.0000 | 0.0000 | 0.0000 | 899.430 | 908.697 | 934.901 | -1817.39 | 8 |
| LAP-HGT | 0.0000 | 0.0000 | 0.0000 | 899.282 | 907.948 | 940.703 | -1815.90 | 10 |
| T-HGED | 0.0000 | 0.0000 | 0.0001 | 896.748 | 906.668 | 939.424 | -1813.34 | 10 |
| T-HGT | 0.0000 | 0.0000 | 0.0000 | 896.220 | 905.583 | 941.614 | -1811.17 | 11 |
| N-HGED | 0.0000 | 0.0000 | 0.0000 | 895.674 | 905.910 | 935.390 | -1811.82 | 9 |
| N-HGT | 0.0000 | 0.0000 | 0.0000 | 894.584 | 902.635 | 935.390 | -1805.27 | 10 |
| GT-HT | 0.0000 | 0.0000 | 0.0000 | 892.498 | 897.006 | 933.037 | -1794.01 | 11 |
| GED-HT | 0.0000 | 0.0000 | 0.0000 | 891.789 | 900.253 | 933.009 | -1800.51 | 10 |
| LAP-HT | 0.0000 | 0.0000 | 0.0000 | 890.543 | 898.370 | 927.850 | -1796.74 | 9 |
| T-HT | 0.0000 | 0.0000 | 0.0000 | 888.969 | 893.926 | 926.681 | -1787.85 | 10 |
| N-HT | 0.0000 | 0.0000 | 0.0000 | 887.846 | 895.842 | 925.322 | -1791.68 | 9 |
| GT-HN | 0.0000 | 0.0000 | 0.0000 | 883.914 | 890.347 | 923.103 | -1780.69 | 10 |
| GED-HN | 0.0000 | 0.0000 | 0.0000 | 882.752 | 893.071 | 922.551 | -1786.14 | 9 |
| T-HN | 0.0000 | 0.0000 | 0.0000 | 880.556 | 889.416 | 918.896 | -1778.83 | 9 |
| LAP-HN | 0.0000 | 0.0000 | 0.0000 | 880.475 | 889.989 | 916.193 | -1779.98 | 8 |
| N-HN | 0.0000 | 0.0000 | 0.0000 | 877.979 | 886.974 | 913.179 | -1773.95 | 8 |

Note: $w_1$ is model weight given by the posterior model probability $P(M_i|y)$ based on the uniform prior; $w_2$ model weight given by the posterior model probability $P(M_i|y)$ based on the OS prior; $w_3$ is model weight based on ML results (against other models); the models are sorted based on Bayesian results under the uniform prior ($w_1$) with GT-GB2 always at the top; $\ln L$ denotes value of the log-likelihood function at maximum, $\ln p(y|M_i)$ represents the so-called *marginal likelihood* or *marginal data density* (MDD), calculated by Laplace approximation at the posterior mode (with Hessian matrix numerically evaluated at the mode); $\ln p_{ML}(y|M_i)$ are ML results calculated based on $\ln p_{ML}(y|M_i) \approx -BIC_i/2$ where $BIC_i = k \ln T - 2 \ln L_i$. We consider two variants of prior model probabilities $p(M_i)$, that is: equal ($p(M_i) \propto c$) labelled 'uniform prior' and $p(M_i) \propto 2^{-k}$ labeled 'OS prior'; with the latter expressing prior preference towards simpler models; $k$ is the number of parameters; model labels are described in Table 1.

Table 3. WHO: posterior means of stochastic parameters across models and model averaging

| Model label | $\sigma_u$ | $\sigma_v$ | $\tau$ | $\psi_u$ | $\psi_v$ | $\nu_u$ | $\nu_v$ |
|---|---|---|---|---|---|---|---|
| GT-GB2 | 0.250 | 0.024 | 0.186 | 32.266 | 1.1684 | 4.27345 | 36.488 |
| LAP-GG | 0.260 | 0.024 | 0.179 | 29.614 | (1) | (inf) | (inf) |
| TLAP-GB2 | 0.251 | 0.023 | 0.178 | 32.726 | (1) | 38.6188 | 42.674 |
| GT-GG | 0.258 | 0.024 | 0.183 | 29.216 | 1.1685 | (inf) | 36.973 |
| GED-GG | 0.259 | 0.024 | 0.181 | 29.918 | 1.5491 | (inf) | (inf) |
| GED-GB2 | 0.251 | 0.024 | 0.181 | 34.285 | 1.4847 | 39.5416 | (inf) |
| T-GB2 | 0.251 | 0.024 | 0.197 | 33.558 | (2) | 39.2235 | 8.4692 |
| N-GG | 0.257 | 0.029 | 0.215 | 33.756 | (2) | (inf) | (inf) |
| LAP-GAM | 0.113 | 0.021 | 0.532 | (1) | (1) | (inf) | (inf) |
| GT-EXP | 0.070 | 0.014 | (1) | (1) | 0.8657 | (inf) | 34.319 |
| GED-EXP | 0.070 | 0.014 | (1) | (1) | 0.6624 | (inf) | (inf) |
| LAP-EXP | 0.072 | 0.018 | (1) | (1) | (1) | (inf) | (inf) |
| GT-HGT | 0.057 | 0.015 | (1) | 0.816 | 0.7256 | 46.7532 | 34.628 |
| GT-HGED | 0.057 | 0.015 | (1) | 0.758 | 0.7443 | (inf) | 35.434 |
| GED-HGED | 0.057 | 0.015 | (1) | 0.761 | 0.6627 | (inf) | (inf) |
| GED-HGT | 0.057 | 0.015 | (1) | 0.817 | 0.6636 | 42.5418 | (inf) |
| T-EXP | 0.077 | 0.019 | (1) | (1) | (2) | (inf) | 26.247 |
| LAP-HGED | 0.057 | 0.020 | (1) | 0.750 | (1) | (inf) | (inf) |
| N-EX | 0.077 | 0.021 | (1) | (1) | (2) | (inf) | (inf) |
| LAP-HGT | 0.055 | 0.020 | (1) | 0.773 | (1) | 45.5269 | (inf) |
| T-HGED | 0.064 | 0.020 | (1) | 0.822 | (2) | (inf) | 16.313 |
| T-HGT | 0.063 | 0.020 | (1) | 0.867 | (2) | 45.8115 | 15.699 |
| N-HGED | 0.072 | 0.022 | (1) | 0.913 | (2) | (inf) | (inf) |
| N-HGT | 0.070 | 0.022 | (1) | 0.953 | (2) | 47.9557 | (inf) |
| GT-HT | 0.062 | 0.014 | (1) | (2) | 0.7289 | 2.83398 | 35.698 |
| GED-HT | 0.061 | 0.014 | (1) | (2) | 0.5682 | 2.72837 | (inf) |
| LAP-HT | 0.069 | 0.018 | (1) | (2) | (1) | 3.24614 | (inf) |
| T-HT | 0.085 | 0.017 | (1) | (2) | (2) | 4.4612 | 24.239 |
| N-HT | 0.084 | 0.020 | (1) | (2) | (2) | 4.43387 | (inf) |
| GT-HN | 0.122 | 0.007 | (1) | (2) | 1.5942 | (inf) | 23.648 |
| GED-HN | 0.123 | 0.006 | (1) | (2) | 0.8493 | (inf) | (inf) |
| T-HN | 0.122 | 0.010 | (1) | (2) | (2) | (inf) | 13.523 |
| LAP-HN | 0.123 | 0.008 | (1) | (2) | (1) | (inf) | (inf) |
| N-HN | 0.120 | 0.015 | (1) | (2) | (2) | (inf) | (inf) |
| Averaged results (informative prior) | 0.256 | 0.024 | 0.181 | 30.868 | 1.122 | 39.313 | 39.076 |
| Averaged results (OS prior) | 0.257 | 0.024 | 0.181 | 30.868 | 2.013 | 68.795 | 61.445 |

Note: values in brackets are fixed (not estimated) in a particular distribution; 'averaged results' refers to results based on formal Bayesian inference pooling (Bayesian model averaging, BMA); model labels are in Table 1.

Table 4. WHO: posterior probabilities of reducing GT-GB2 based on two prior probabilities

| $p(M_i)$ | $P(\tau = 1|y)$ | $P(\psi_u = 1|y)$ | $P(\psi_u = 2|y)$ | $P(\psi_v = 1|y)$ | $P(\psi_v = 2|y)$ | $P(\nu_u \to \infty|y)$ | $P(\nu_v \to \infty|y)$ |
|---|---|---|---|---|---|---|---|
| uniform | 0 | 0 | 0 | 0.626 | 0.007 | 0.604 | 0.569 |
| OS | 0 | 0 | 0 | 0.790 | 0.005 | 0.774 | 0.726 |

Note: $p(M_i)$ is model prior; $p(M_i) \propto c$ labeled 'uniform'; $p(M_i) \propto 2^{-k}$ labeled 'OS'.

Table 5. WHO: comparison of efficiency scores across models

| Model label | $w_1$ | mean | min | max | range |
|---|---|---|---|---|---|
| GT-GB2 | 0.1320 | 0.958 | 0.749 | 0.995 | 0.246 |
| LAP-GG | 0.4259 | 0.959 | 0.755 | 0.996 | 0.240 |
| TLAP-GB2 | 0.2005 | 0.959 | 0.750 | 0.995 | 0.245 |
| GT-GG | 0.0925 | 0.958 | 0.756 | 0.996 | 0.239 |
| GED-GG | 0.0855 | 0.959 | 0.755 | 0.996 | 0.241 |
| GEF-GB2 | 0.0571 | 0.958 | 0.749 | 0.995 | 0.247 |
| T-GB2 | 0.0061 | 0.956 | 0.750 | 0.996 | 0.246 |
| N-GG | 0.0005 | 0.953 | 0.749 | 0.998 | 0.249 |
| LAP-GAM | 0.0000 | 0.949 | 0.709 | 0.991 | 0.283 |
| GT-EXP | 0.0000 | 0.935 | 0.711 | 0.988 | 0.277 |
| GED-EXP | 0.0000 | 0.935 | 0.710 | 0.988 | 0.278 |
| LAP-EXP | 0.0000 | 0.933 | 0.700 | 0.986 | 0.286 |
| GT-HGT | 0.0000 | 0.939 | 0.709 | 0.988 | 0.279 |
| GT-HGED | 0.0000 | 0.940 | 0.709 | 0.988 | 0.279 |
| GED-HGED | 0.0000 | 0.939 | 0.708 | 0.988 | 0.281 |
| GED-HGT | 0.0000 | 0.939 | 0.708 | 0.988 | 0.280 |
| T-EXP | 0.0000 | 0.929 | 0.699 | 0.990 | 0.292 |
| LAP-HGED | 0.0000 | 0.939 | 0.702 | 0.987 | 0.285 |
| N-EX | 0.0000 | 0.928 | 0.699 | 0.995 | 0.296 |
| LAP-HGT | 0.0000 | 0.939 | 0.702 | 0.987 | 0.285 |
| T-HGED | 0.0000 | 0.935 | 0.704 | 0.989 | 0.285 |
| T-HGT | 0.0000 | 0.935 | 0.704 | 0.989 | 0.285 |
| N-HGED | 0.0000 | 0.931 | 0.700 | 0.995 | 0.294 |
| N-HGT | 0.0000 | 0.931 | 0.700 | 0.995 | 0.295 |
| GT-HT | 0.0000 | 0.935 | 0.711 | 0.986 | 0.275 |
| GED-HT | 0.0000 | 0.913 | 0.691 | 0.993 | 0.302 |
| LAP-HT | 0.0000 | 0.936 | 0.711 | 0.986 | 0.275 |
| T-HT | 0.0000 | 0.931 | 0.697 | 0.985 | 0.288 |
| N-HT | 0.0000 | 0.923 | 0.694 | 0.990 | 0.296 |
| GT-HN | 0.0000 | 0.923 | 0.694 | 0.995 | 0.300 |
| GED-HN | 0.0000 | 0.913 | 0.693 | 0.993 | 0.299 |
| T-HN | 0.0000 | 0.914 | 0.693 | 0.993 | 0.300 |
| LAP-HN | 0.0000 | 0.914 | 0.691 | 0.992 | 0.300 |
| N-HN | 0.0000 | 0.915 | 0.694 | 0.996 | 0.302 |
| Averaged results (informative prior) | - | 0.959 | 0.753 | 0.995 | 0.242 |
| Averaged results (OS prior) | - | 0.959 | 0.755 | 0.996 | 0.241 |

Note: $w_1$ is model weight based on uniform prior; see notes for Table 2.

Table 6. Spanish dairy farms: model comparison results

| Model label | $w_1$ | $w_2$ | $w_3$ | $\ln p(y\|M_i)$ | $\ln p_{ML}(y\|M_i)$ | $\ln L$ | BIC | $k$ |
|---|---|---|---|---|---|---|---|---|
| GT-GB2 | 0.2013 | 0.0174 | 0.0000 | 810.518 | 788.053 | 872.016 | -1576.11 | 23 |
| LAP-HN | 0.1967 | 0.5446 | 0.8824 | 810.495 | 805.100 | 870.810 | -1610.20 | 18 |
| LAP-HT | 0.1073 | 0.1485 | 0.0236 | 809.889 | 801.478 | 870.838 | -1602.96 | 19 |
| GT-N | 0.0886 | 0.0613 | 0.0011 | 809.697 | 798.380 | 871.391 | -1596.76 | 20 |
| GT-GG | 0.0839 | 0.0145 | 0.0000 | 809.642 | 791.486 | 871.799 | -1582.97 | 22 |
| GT-HT | 0.0737 | 0.0255 | 0.0000 | 809.513 | 794.824 | 871.486 | -1589.65 | 21 |
| T-HT | 0.0512 | 0.0354 | 0.0000 | 809.149 | 791.118 | 871.485 | -1596.95 | 20 |
| T-HN | 0.0462 | 0.0640 | 0.0012 | 809.047 | 798.473 | 871.389 | -1604.06 | 19 |
| GED-HN | 0.0249 | 0.0344 | 0.0230 | 808.427 | 801.454 | 870.815 | -1602.91 | 19 |
| TLAP-GB2 | 0.0194 | 0.0034 | 0.0007 | 808.180 | 798.016 | 871.430 | -1582.24 | 22 |
| T-GB2 | 0.0192 | 0.0033 | 0.0011 | 808.167 | 798.450 | 871.793 | -1582.96 | 22 |
| GED-HT | 0.0141 | 0.0098 | 0.0006 | 807.861 | 797.827 | 870.839 | -1595.65 | 20 |
| LAP-GG | 0.0118 | 0.0082 | 0.0008 | 807.680 | 798.099 | 871.111 | -1596.20 | 20 |
| GT-HGT | 0.0110 | 0.0019 | 0.0000 | 807.610 | 791.181 | 871.494 | -1582.36 | 22 |
| LAP-HGT | 0.0103 | 0.0071 | 0.0008 | 807.544 | 798.046 | 871.057 | -1596.09 | 20 |
| GED-GB2 | 0.0087 | 0.0015 | 0.0000 | 807.379 | 791.022 | 871.335 | -1582.04 | 22 |
| GED-GG | 0.0078 | 0.0027 | 0.0000 | 807.271 | 794.447 | 871.109 | -1588.89 | 21 |
| T-HGT | 0.0051 | 0.0018 | 0.0409 | 806.836 | 802.028 | 871.491 | -1589.66 | 21 |
| GED-HGT | 0.0044 | 0.0015 | 0.0000 | 806.692 | 794.428 | 871.090 | -1588.86 | 21 |
| LAP-HGED | 0.0042 | 0.0058 | 0.0229 | 806.649 | 801.450 | 870.811 | -1602.90 | 19 |
| GT-HGED | 0.0034 | 0.0012 | 0.0000 | 806.449 | 794.813 | 871.475 | -1589.63 | 21 |
| T-EXP | 0.0020 | 0.0028 | 0.0000 | 805.925 | 791.480 | 867.377 | -1596.03 | 19 |
| T-HGED | 0.0020 | 0.0014 | 0.0000 | 805.919 | 794.829 | 871.461 | -1596.90 | 20 |
| GT-EXP | 0.0017 | 0.0012 | 0.0001 | 805.768 | 795.683 | 868.694 | -1591.37 | 20 |
| GED-HGED | 0.0009 | 0.0006 | 0.0006 | 805.066 | 797.805 | 870.816 | -1595.61 | 20 |
| GED-EXP | 0.0001 | 0.0001 | 0.0001 | 802.312 | 795.458 | 864.819 | -1590.92 | 19 |
| N-GG | 0.0000 | 0.0000 | 0.0000 | 801.683 | 790.378 | 863.390 | -1580.76 | 20 |
| N-EX | 0.0000 | 0.0000 | 0.0000 | 799.560 | 794.287 | 859.997 | -1588.57 | 18 |
| N-HT | 0.0000 | 0.0000 | 0.0000 | 798.747 | 790.087 | 859.447 | -1580.17 | 19 |
| LAP-EXP | 0.0000 | 0.0000 | 0.0000 | 797.613 | 791.718 | 857.428 | -1583.44 | 18 |
| N-HN | 0.0000 | 0.0000 | 0.0000 | 796.919 | 791.545 | 857.256 | -1583.09 | 18 |
| N-HGT | 0.0000 | 0.0000 | 0.0000 | 796.380 | 787.460 | 860.472 | -1574.92 | 20 |
| N-HGED | 0.0000 | 0.0000 | 0.0000 | 796.331 | 791.111 | 860.472 | -1582.22 | 19 |
| LAP-GAM | 0.0000 | 0.0000 | 0.0000 | 795.590 | 788.707 | 858.068 | -1577.41 | 19 |

See notes for Table 2.

Table 7. Spanish dairy farms: posterior probabilities of reducing GT-GB2 based on two prior probabilities

| $p(M_i)$ | $P(\tau=1\|y)$ | $P(\psi_u=1\|y)$ | $P(\psi_u=2\|y)$ | $P(\psi_v=1\|y)$ | $P(\psi_v=2\|y)$ | $P(v_u \to \infty\|y)$ | $P(v_v \to \infty\|y)$ |
|---|---|---|---|---|---|---|---|
| uniform | 0.648 | 0.004 | 0.603 | 0.350 | 0.126 | 0.474 | 0.391 |
| OS | 0.949 | 0.004 | 0.924 | 0.718 | 0.109 | 0.743 | 0.765 |

See notes for Table 4.

Figure 1. WHO: 2D heatmaps of marginal posterior densities for $\sigma_u$, $\sigma_v$, $\tau$ in GT-GB2

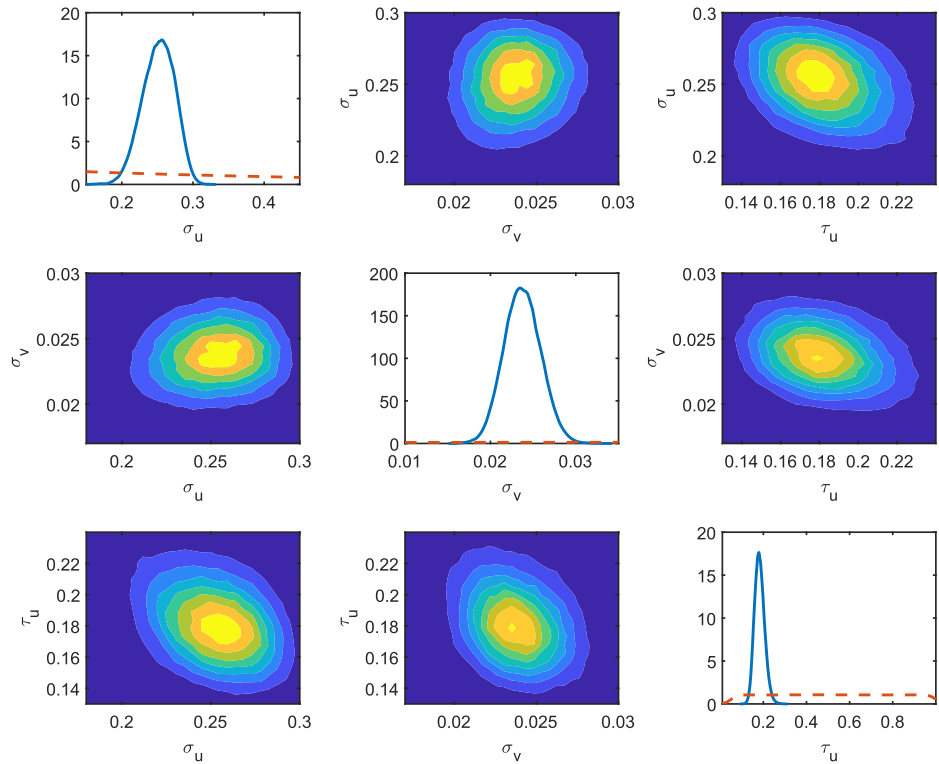

Note: parameters' 1D marginal posterior densities are on the diagonal (dashed lines represent their marginal prior densities); full heatmap matrix can be viewed in Appendix 2.

Figure 2. WHO: posterior density of average efficiency under uniform and OS priors

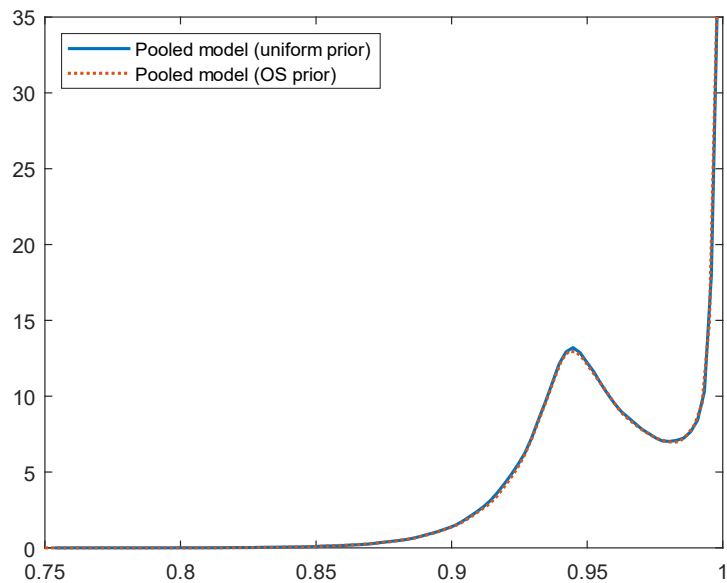

Note: pooled/averaged efficiency densities under uniform and OS priors overlap.

Figure 3. Spanish dairy farms: 2D heatmaps of marginal posteriors densities for $\sigma_u$, $\sigma_v$, $\tau$ in GT-GB2

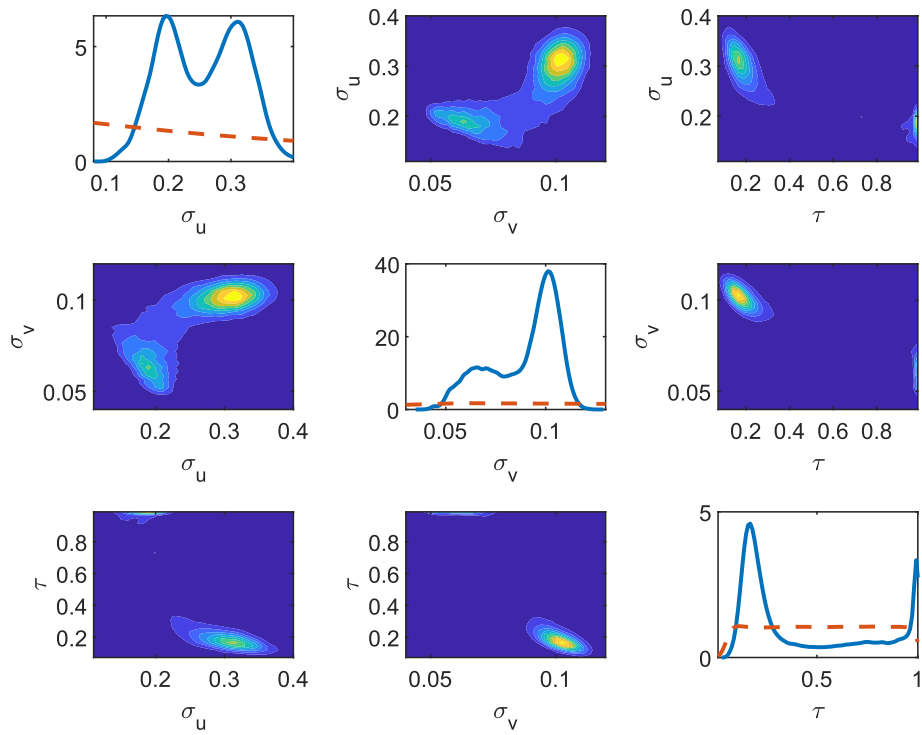

Note: parameters' 1D marginal posterior densities are on the diagonal (dashed lines represent marginal prior densities); full heatmap matrix can be viewed in Appendix 2.

Figure 4. Spanish dairy farms: posterior density of average efficiency under uniform and OS priors

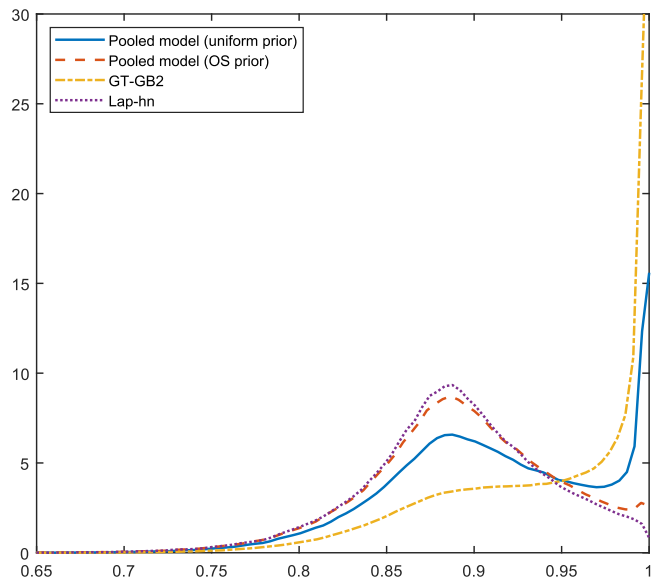

# Appendices (online supplementary materials)

## Appendix 1. Details regarding model specification and technical aspects of inference.

Statistical inference in the GT-GB2 SF model relies upon adequate evaluation of the log-likelihood. This in turn requires reliable evaluation of the integral in (12) at different points of the parameter space. As the integral is one-dimensional, we suggest the use of traditional, non-stochastic methods (see, e.g., Shampine, 2008). Our experience is, however, that standard numerical procedures require some fine-tuning, like specifying relevant integration waypoints, to obtain satisfactory precision. As the integrand is defined as a product of GB2 and truncated GT densities, it is likely to be multimodal in some cases (depending on location of the truncation point). It is therefore vital to impose knots at points that correspond to (at least approximate) location of the modes, taking into consideration possible contrast between $\sigma_u$ and $\sigma_v$ as well. Our experience indicates that such tuned procedures provide satisfactory precision of the likelihood evaluation.

Available methods of multi-core parallelization, e.g. MATLAB, provide an efficient way of reducing the computational time as each of $T$ integrals can be computed independently, not to mention that one can run multiple estimation procedures simultaneously. Our experience is that by optimizing parallelization and vectorization it is possible to obtain acceptable computational time for full Bayesian estimations using fairly standard single-processor workstations (e.g., we have used MATLAB 2019a release with Parallel Computing Toolbox on a PC with 64-thread AMD Threadripper 3970X CPU and 64GB RAM; this allowed us to run up to 8 simulations at once, each having assigned from 6 to 8 workers with up to 0.5 million iterations per model; the total RAM usage would not exceed 45GB and the results were available within a day).

A complete Stochastic Frontier Analysis obviously requires inference on latent variables that represent individual (observation or object-specific) inefficiency terms (u's). Moreover, a nonlinear transformation of u's, having the form of $r = \exp(-u)$, provides the so-called efficiency factors (r). Note that within the Bayesian approach it is possible to draw u's conditionally on the statistical parameters and the data. Within the GT-GB2 model the terms should be drawn from a distribution that is nonstandard. However, a MH-step can be used to obtain the sample. Obviously, this requires an efficient proposal. The target distribution corresponds to the integrand function mentioned in Section 3 (being a product of densities, see (2) and (12)):

$$p(u_t|\theta, y) \propto \left[\frac{1}{v_v}\left(\frac{|\ln y_t - \ln g(x_t;\beta) + u_t|}{\sigma_v}\right)^{\psi_v} + 1\right]^{-\frac{(1+v_v)}{\psi_v}} \left(\frac{u_t}{\sigma_u}\right)^{\tau-1} \left[\frac{1}{v_u}\left(\frac{u_t}{\sigma_u}\right)^{\psi_u} + 1\right]^{-\frac{(\tau+v_u)}{\psi_u}} \quad (A1.1)$$

We suggest to use of a mixture of the respective variables, GB2 and truncated GT, with scale parameters $\sigma_u$ and $\sigma_v$ multiplied by some constant > 1, e.g. 1.5, as a proposal. It is feasible to sample from such a mixture and it is likely to approximate the relevant location of modes. In order to sample from the GB2 distribution, one might make use of the following procedure: firstly, draw $X_1 \sim \chi^2(2\tau/\psi)$ and $X_2 \sim \chi^2(2v/\psi)$; compute $Z = X_1/X_2$, then $Y = Z^{1/\psi}$, and $W = \sigma v^{1/\psi} Y$, where $W$ is distributed as $f_{GB2}(.;\sigma, v, \psi, \tau)$. The procedure can be modified using the relationship between GT and GB2 making it possible to sample from the GT distribution as well. Our experience has shown that such a mixture-like proposal is sufficient to draw inefficiency terms. Additionally, if the truncation point is far in tails of the GT distribution, the GT mixture component can be either omitted or replaced by non-truncated GT in order to make the computation simpler (avoiding the problem of drawing from truncated densities with truncation point located in upper tail).

Obviously, as an alternative one might transform the target density to some interval (e.g. by drawing efficiency indices, denoted by r) with some form of Beta-type proposal. Note that all the inefficiency terms can be drawn individually. Univariate sampling from unknown density is relatively simple, and the sampling procedure might be parallelized. Moreover, contrary to the usual Gibbs sampling, draws of u's (or their functions) are not required in order to draw $\theta$. Hence, imperfect procedure for sampling u's is not going to deteriorate inference on model parameters. Moreover, one might draw only some of u's, representing the objects of interest, or use alternative methods to estimate quantile characteristics of u's. Generally, although u's need to be sampled in a non-standard way (and the target density should use subsequent sampled values of $\theta's$), in general it is unlikely to cause significant practical problems.

Finally, note that it is possible to conduct approximate inference on inefficiency terms assuming that the posterior for $\theta's$ is point mass. Hence, it is sufficient to plot and analyze the conditional distribution (A1.1) plugging in point estimates of $\boldsymbol{\theta}'s$ (posterior mean/mode or MLEs). Such plots (after normalization) can be averaged across models using calculated model weights (e.g., via posterior model probabilities), resulting in a density estimates of object-specific u's obtained by model averaging.

Taking the above into consideration it is possible to use the following approximate inference strategy: (1) obtain point estimates of parameters (MLEs/maximum posterior), and posterior model probabilities (based on Laplace approximation or BICs), (2) compute approximate density estimates of inefficiency terms (see above) (3) average the density estimates across models using posterior model probabilities as weights. Consequently, one may obtain (approximate) density estimates of inefficiency terms by model averaging, without the use of computationally intensive MCMC-type methods. Obviously, a full Bayesian MCMC procedure is likely to be more reliable (though more costly in terms of computational time).

In order to verify performance of full MCMC procedures outlined in the paper, we have compared the performance of our full Bayesian algorithm with results obtained using well-known Gibbs samplers (for normal half-normal and normal-exponential models) and have found practically identical results in terms of inference on parameters and inefficiencies; though of course there were considerable differences in terms of computation time. The estimates were also checked against a popular program for estimating SF models (LIMDEP 11) with, again, practically identical results in terms of inference on parameters and inefficiencies.

Bayesian estimation requires formulation of prior beliefs on possible values of model parameters. However, nowadays priors are often motivated by properties of numerical methods necessary to sample from posteriors. Our approach makes it possible to overcome this requirement and thus to conduct sensitivity analyses. In the empirical part of the paper we have assumed that all the model parameters in GT-GB2 are a priori independent and follow proper priors. Moreover, we have assumed that a priori:

$$\tau \sim U(0.05,1), \qquad \sigma_v, \sigma_u \sim GG(1,1,1), \qquad (\psi_v - 1) \sim GG(2,1,1),$$
$$(\psi_u - 1) \sim GG(30,1,1), \ (\nu_u - 2) \sim GG(30,1,1), \ (\nu_v - 2) \sim GG(30,1,1) \qquad (A1.2)$$
$$\beta \sim t_k(3, 0_k, 10 I_k)$$

where $U(a,b)$ is the uniform distribution between $a$ and $b$, $t_k(\nu, \mu, A)$ is the multivariate $t$ distribution with degrees of freedom $\nu$, $k$x1 location vector $\mu$ and $k$x$k$ precision matrix $A$, $GG(\sigma, \tau, \psi)$ is the generalized gamma distribution with parameters according to Equation (9) in Section 3. These assumptions are motivated by authors experience in the field; in particular the goal is to make them as simple and intuitively weakly informative as possible. We use different priors for $\psi_v$ and $\psi_u$, since allowing for $\tau < 1$ implies that the role of the two parameters is somewhat different. Formal elicitation of well-justified priors for the GT-GB2 specification is left for further research.

For the purpose of efficient numerical evaluation, we make use of some fairly standard transformations of parameters (where necessary) to the unconstrained real values (solely for the purpose of computation). This of course requires transformation of priors as well. However, we find it beneficial as the optimization algorithms, MCMC samplers as well as the Laplace-type approximations are likely to work better in an unconstrained parameter space.

As our formulation nests a number of well-established models (including the most popular ones) it allows for the so-called coherence analysis (Makieła and Mazur, 2020). The analysis provides a way to verify whether prior beliefs that are assumed in specific reduced models can be shown to arise from prior beliefs formulated in a general case. Consequently, within our framework it is possible to re-evaluate standard priors used so far in Bayesian SFA. This is an important issue from the model averaging viewpoint. It is well-known that Bayesian model averaging (BMA) depends crucially upon the so-called Bayes factors. The latter in turn are sensitive to prior assumptions. Consequently, checks for prior sensitivity and coherence are particularly important if one aims to properly deal with model specification uncertainty.

# Appendix 2. Additional tables and figures

Table A1. WHO: posterior means of translog function parameters

| Model label | $\beta_0$ | $\beta_1$ | $\beta_2$ | $\beta_3$ | $\beta_4$ | $\beta_5$ |
|---|---|---|---|---|---|---|
| GT-GB2 | 0.059 | 0.090 | 0.047 | -0.006 | 0.050 | -0.039 |
| LAP-GG | 0.058 | 0.091 | 0.046 | -0.006 | 0.051 | -0.039 |
| TLAP-GB2 | 0.059 | 0.092 | 0.045 | -0.006 | 0.052 | -0.039 |
| GT-GG | 0.058 | 0.090 | 0.047 | -0.006 | 0.050 | -0.039 |
| GED-GG | 0.058 | 0.090 | 0.047 | -0.006 | 0.050 | -0.039 |
| GEF-GB2 | 0.058 | 0.090 | 0.047 | -0.006 | 0.050 | -0.039 |
| T-GB2 | 0.059 | 0.088 | 0.050 | -0.005 | 0.048 | -0.038 |
| N-GG | 0.058 | 0.080 | 0.058 | 0.000 | 0.047 | -0.043 |
| LAP-GAM | 0.072 | 0.088 | 0.050 | -0.007 | 0.039 | -0.034 |
| GT-EXP | 0.086 | 0.086 | 0.054 | -0.007 | 0.037 | -0.031 |
| GED-EXP | 0.086 | 0.087 | 0.054 | -0.007 | 0.036 | -0.030 |
| LAP-EXP | 0.086 | 0.084 | 0.056 | -0.005 | 0.035 | -0.034 |
| GT-HGT | 0.083 | 0.089 | 0.051 | -0.008 | 0.039 | -0.030 |
| GT-HGED | 0.082 | 0.089 | 0.051 | -0.008 | 0.038 | -0.030 |
| GED-HGED | 0.082 | 0.089 | 0.052 | -0.008 | 0.037 | -0.029 |
| GED-HGT | 0.083 | 0.089 | 0.051 | -0.008 | 0.038 | -0.030 |
| T-EXP | 0.087 | 0.078 | 0.061 | 0.000 | 0.037 | -0.042 |
| LAP-HGED | 0.081 | 0.086 | 0.053 | -0.006 | 0.036 | -0.033 |
| N-EX | 0.086 | 0.077 | 0.062 | 0.001 | 0.040 | -0.043 |
| LAP-HGT | 0.081 | 0.087 | 0.053 | -0.006 | 0.036 | -0.033 |
| T-HGED | 0.082 | 0.082 | 0.056 | -0.003 | 0.038 | -0.038 |
| T-HGT | 0.082 | 0.082 | 0.056 | -0.003 | 0.038 | -0.038 |
| N-HGED | 0.084 | 0.078 | 0.061 | 0.000 | 0.039 | -0.042 |
| N-HGT | 0.083 | 0.078 | 0.061 | 0.000 | 0.039 | -0.042 |
| GT-HT | 0.087 | 0.088 | 0.053 | -0.007 | 0.038 | -0.030 |
| GED-HT | 0.086 | 0.089 | 0.053 | -0.008 | 0.037 | -0.029 |
| LAP-HT | 0.088 | 0.084 | 0.058 | -0.004 | 0.035 | -0.035 |
| T-HT | 0.092 | 0.076 | 0.064 | 0.000 | 0.040 | -0.043 |
| N-HT | 0.091 | 0.076 | 0.065 | 0.001 | 0.041 | -0.043 |
| GT-HN | 0.100 | 0.070 | 0.077 | 0.001 | 0.049 | -0.045 |
| GED-HN | 0.101 | 0.069 | 0.077 | 0.001 | 0.049 | -0.045 |
| T-HN | 0.099 | 0.070 | 0.076 | 0.001 | 0.049 | -0.045 |
| LAP-HN | 0.099 | 0.070 | 0.076 | 0.001 | 0.049 | -0.046 |
| N-HN | 0.096 | 0.072 | 0.074 | 0.001 | 0.047 | -0.043 |
| Averaged results (informative prior) | 0.058 | 0.091 | 0.046 | -0.006 | 0.051 | -0.039 |
| Averaged results (OS prior) | 0.058 | 0.091 | 0.046 | -0.006 | 0.051 | -0.039 |

Note: variables have been mean-corrected prior to estimation; thus $\beta_1$ and $\beta_2$ represent factor elasticities at sample mean.

Table A2. WHO: posterior standard deviation of translog function parameters

| Model label | $\beta_0$ | $\beta_1$ | $\beta_2$ | $\beta_3$ | $\beta_4$ | $\beta_5$ |
|---|---|---|---|---|---|---|
| GT-GB2 | 0.0045 | 0.0039 | 0.0060 | 0.0022 | 0.0109 | 0.0065 |
| LAP-GG | 0.0043 | 0.0034 | 0.0052 | 0.0019 | 0.0109 | 0.0059 |
| TLAP-GB2 | 0.0042 | 0.0033 | 0.0052 | 0.0019 | 0.0105 | 0.0058 |
| GT-GG | 0.0041 | 0.0039 | 0.0059 | 0.0022 | 0.0113 | 0.0065 |
| GED-GG | 0.0043 | 0.0041 | 0.0064 | 0.0022 | 0.0113 | 0.0063 |
| GEF-GB2 | 0.0042 | 0.0040 | 0.0058 | 0.0022 | 0.0109 | 0.0062 |
| T-GB2 | 0.0036 | 0.0043 | 0.0068 | 0.0027 | 0.0104 | 0.0069 |
| N-GG | 0.0038 | 0.0028 | 0.0054 | 0.0021 | 0.0108 | 0.0071 |
| LAP-GAM | 0.0050 | 0.0033 | 0.0059 | 0.0020 | 0.0107 | 0.0067 |
| GT-EXP | 0.0030 | 0.0042 | 0.0058 | 0.0033 | 0.0111 | 0.0084 |
| GED-EXP | 0.0030 | 0.0036 | 0.0055 | 0.0029 | 0.0107 | 0.0075 |
| LAP-EXP | 0.0033 | 0.0032 | 0.0058 | 0.0025 | 0.0107 | 0.0078 |
| GT-HGT | 0.0041 | 0.0037 | 0.0078 | 0.0024 | 0.0107 | 0.0068 |
| GT-HGED | 0.0034 | 0.0032 | 0.0058 | 0.0023 | 0.0106 | 0.0068 |
| GED-HGED | 0.0036 | 0.0031 | 0.0053 | 0.0022 | 0.0105 | 0.0064 |
| GED-HGT | 0.0034 | 0.0032 | 0.0055 | 0.0024 | 0.0104 | 0.0067 |
| T-EXP | 0.0036 | 0.0037 | 0.0061 | 0.0026 | 0.0119 | 0.0083 |
| LAP-HGED | 0.0038 | 0.0030 | 0.0055 | 0.0022 | 0.0104 | 0.0069 |
| N-EX | 0.0036 | 0.0032 | 0.0060 | 0.0021 | 0.0118 | 0.0079 |
| LAP-HGT | 0.0040 | 0.0030 | 0.0055 | 0.0022 | 0.0105 | 0.0069 |
| T-HGED | 0.0049 | 0.0055 | 0.0072 | 0.0037 | 0.0115 | 0.0089 |
| T-HGT | 0.0049 | 0.0056 | 0.0072 | 0.0038 | 0.0116 | 0.0089 |
| N-HGED | 0.0052 | 0.0033 | 0.0071 | 0.0021 | 0.0116 | 0.0078 |
| N-HGT | 0.0045 | 0.0033 | 0.0066 | 0.0021 | 0.0115 | 0.0077 |
| GT-HT | 0.0036 | 0.0045 | 0.0069 | 0.0032 | 0.0110 | 0.0077 |
| GED-HT | 0.0032 | 0.0034 | 0.0058 | 0.0025 | 0.0108 | 0.0068 |
| LAP-HT | 0.0049 | 0.0051 | 0.0071 | 0.0031 | 0.0112 | 0.0083 |
| T-HT | 0.0044 | 0.0042 | 0.0069 | 0.0027 | 0.0120 | 0.0086 |
| N-HT | 0.0041 | 0.0034 | 0.0065 | 0.0023 | 0.0120 | 0.0081 |
| GT-HN | 0.0052 | 0.0039 | 0.0074 | 0.0027 | 0.0122 | 0.0088 |
| GED-HN | 0.0046 | 0.0036 | 0.0070 | 0.0026 | 0.0122 | 0.0085 |
| T-HN | 0.0040 | 0.0029 | 0.0069 | 0.0024 | 0.0117 | 0.0083 |
| LAP-HN | 0.0034 | 0.0026 | 0.0071 | 0.0024 | 0.0122 | 0.0084 |
| N-HN | 0.0034 | 0.0030 | 0.0070 | 0.0024 | 0.0117 | 0.0083 |

Table A3. WHO: correlations between efficiency rankings across different models

| | gt-gb2 | lap-gg | tlap-gb2 | gt-gg | ged-gg | gef-gb2 | t-gb2 | n-gg | lap-gam | gt-exp | ged-exp | lap-exp | gt-hgt | gt-hged | ged-hged | ged-hgt | t-exp | lap-hged | n-ex | lap-hgt | t-hged | t-hgt | n-hged | n-hgt | gt-ht | ged-ht | lap-ht | t-ht | n-ht | gt-hn | ged-hn | t-hn | lap-hn | n-hn |
|---|---|---|---|---|---|---|---|---|---|---|---|---|---|---|---|---|---|---|---|---|---|---|---|---|---|---|---|---|---|---|---|---|---|---|
| gt-gb2 | 1.000 | 0.999 | 0.999 | 0.973 | 1.000 | 1.000 | 0.987 | 0.987 | 0.986 | 0.986 | 0.996 | 0.986 | 0.987 | 0.999 | 0.999 | 1.000 | 0.973 | 0.999 | 0.999 | 0.999 | 0.974 | 0.998 | 0.990 | 0.992 | 0.992 | 0.979 | 0.989 | 0.986 | 0.992 | 0.989 | 0.997 | 0.997 | 0.975 | 0.989 |
| lap-gg | 0.999 | 1.000 | 1.000 | 0.966 | 1.000 | 0.999 | 0.991 | 0.990 | 0.990 | 0.990 | 0.998 | 0.986 | 0.990 | 1.000 | 1.000 | 0.999 | 0.966 | 0.997 | 0.999 | 0.999 | 0.967 | 0.997 | 0.986 | 0.988 | 0.988 | 0.972 | 0.984 | 0.990 | 0.988 | 0.991 | 0.995 | 0.995 | 0.968 | 0.984 |
| tlap-gb2 | 0.999 | 1.000 | 1.000 | 0.966 | 1.000 | 0.999 | 0.991 | 0.990 | 0.990 | 0.990 | 0.998 | 0.986 | 0.990 | 1.000 | 1.000 | 1.000 | 0.966 | 0.997 | 0.999 | 0.999 | 0.967 | 0.997 | 0.986 | 0.988 | 0.988 | 0.973 | 0.985 | 0.990 | 0.988 | 0.991 | 0.995 | 0.996 | 0.968 | 0.985 |
| gt-gg | 0.973 | 0.966 | 0.966 | 1.000 | 0.967 | 0.975 | 0.945 | 0.944 | 0.944 | 0.942 | 0.959 | 0.967 | 0.944 | 0.967 | 0.966 | 0.970 | 1.000 | 0.980 | 0.970 | 0.969 | 1.000 | 0.980 | 0.992 | 0.989 | 0.989 | 0.998 | 0.994 | 0.940 | 0.991 | 0.954 | 0.982 | 0.982 | 1.000 | 0.994 |
| ged-gg | 1.000 | 1.000 | 1.000 | 0.967 | 1.000 | 0.999 | 0.987 | 0.987 | 0.987 | 0.986 | 0.997 | 0.983 | 0.987 | 1.000 | 1.000 | 1.000 | 0.967 | 0.997 | 0.998 | 0.998 | 0.969 | 0.997 | 0.986 | 0.988 | 0.988 | 0.974 | 0.985 | 0.986 | 0.988 | 0.988 | 0.995 | 0.995 | 0.970 | 0.985 |
| gef-gb2 | 1.000 | 0.999 | 0.999 | 0.975 | 0.999 | 1.000 | 0.987 | 0.986 | 0.986 | 0.985 | 0.996 | 0.987 | 0.986 | 0.999 | 0.999 | 1.000 | 0.975 | 0.999 | 0.999 | 0.999 | 0.976 | 0.998 | 0.991 | 0.992 | 0.993 | 0.981 | 0.990 | 0.985 | 0.993 | 0.989 | 0.997 | 0.998 | 0.977 | 0.990 |
| t-gb2 | 0.987 | 0.991 | 0.991 | 0.945 | 0.987 | 0.987 | 1.000 | 1.000 | 1.000 | 1.000 | 0.996 | 0.992 | 1.000 | 0.991 | 0.991 | 0.987 | 0.945 | 0.986 | 0.992 | 0.992 | 0.947 | 0.985 | 0.974 | 0.977 | 0.978 | 0.954 | 0.970 | 1.000 | 0.975 | 0.999 | 0.986 | 0.986 | 0.948 | 0.969 |
| n-gg | 0.987 | 0.990 | 0.990 | 0.944 | 0.987 | 0.986 | 1.000 | 1.000 | 1.000 | 1.000 | 0.995 | 0.992 | 1.000 | 0.990 | 0.990 | 0.986 | 0.944 | 0.985 | 0.991 | 0.991 | 0.946 | 0.984 | 0.973 | 0.977 | 0.977 | 0.953 | 0.969 | 1.000 | 0.974 | 0.999 | 0.985 | 0.985 | 0.947 | 0.968 |
| lap-gam | 0.986 | 0.990 | 0.990 | 0.944 | 0.987 | 0.986 | 1.000 | 1.000 | 1.000 | 1.000 | 0.995 | 0.992 | 1.000 | 0.990 | 0.990 | 0.986 | 0.944 | 0.985 | 0.991 | 0.991 | 0.946 | 0.984 | 0.973 | 0.977 | 0.977 | 0.953 | 0.969 | 1.000 | 0.974 | 0.999 | 0.985 | 0.985 | 0.947 | 0.968 |
| gt-exp | 0.986 | 0.990 | 0.990 | 0.942 | 0.986 | 0.985 | 1.000 | 1.000 | 1.000 | 1.000 | 0.995 | 0.991 | 1.000 | 0.990 | 0.990 | 0.986 | 0.942 | 0.984 | 0.991 | 0.991 | 0.944 | 0.983 | 0.972 | 0.975 | 0.976 | 0.951 | 0.968 | 1.000 | 0.973 | 0.999 | 0.984 | 0.984 | 0.945 | 0.967 |
| ged-exp | 0.996 | 0.998 | 0.998 | 0.959 | 0.997 | 0.996 | 0.996 | 0.995 | 0.995 | 0.995 | 1.000 | 0.990 | 0.996 | 0.998 | 0.998 | 0.996 | 0.959 | 0.995 | 0.999 | 0.999 | 0.961 | 0.995 | 0.984 | 0.986 | 0.987 | 0.967 | 0.981 | 0.995 | 0.985 | 0.995 | 0.994 | 0.994 | 0.962 | 0.981 |
| lap-exp | 0.986 | 0.986 | 0.986 | 0.967 | 0.983 | 0.987 | 0.992 | 0.992 | 0.992 | 0.991 | 0.990 | 1.000 | 0.992 | 0.986 | 0.986 | 0.984 | 0.967 | 0.989 | 0.990 | 0.990 | 0.969 | 0.988 | 0.988 | 0.990 | 0.991 | 0.974 | 0.985 | 0.989 | 0.988 | 0.996 | 0.992 | 0.992 | 0.969 | 0.984 |
| gt-hgt | 0.987 | 0.990 | 0.990 | 0.944 | 0.987 | 0.986 | 1.000 | 1.000 | 1.000 | 1.000 | 0.996 | 0.992 | 1.000 | 0.990 | 0.990 | 0.987 | 0.945 | 0.985 | 0.992 | 0.992 | 0.946 | 0.984 | 0.973 | 0.977 | 0.977 | 0.953 | 0.970 | 1.000 | 0.974 | 0.999 | 0.985 | 0.986 | 0.947 | 0.968 |
| gt-hged | 0.999 | 1.000 | 1.000 | 0.967 | 1.000 | 0.999 | 0.991 | 0.990 | 0.990 | 0.990 | 0.998 | 0.986 | 0.990 | 1.000 | 1.000 | 1.000 | 0.967 | 0.998 | 0.999 | 0.999 | 0.968 | 0.997 | 0.987 | 0.989 | 0.989 | 0.974 | 0.985 | 0.990 | 0.989 | 0.991 | 0.996 | 0.996 | 0.969 | 0.985 |
| ged-hged | 0.999 | 1.000 | 1.000 | 0.966 | 1.000 | 0.999 | 0.991 | 0.990 | 0.990 | 0.990 | 0.998 | 0.986 | 0.990 | 1.000 | 1.000 | 0.999 | 0.966 | 0.997 | 0.999 | 0.999 | 0.967 | 0.996 | 0.986 | 0.988 | 0.988 | 0.972 | 0.984 | 0.990 | 0.988 | 0.991 | 0.995 | 0.995 | 0.968 | 0.984 |
| ged-hgt | 1.000 | 0.999 | 1.000 | 0.970 | 1.000 | 1.000 | 0.987 | 0.986 | 0.986 | 0.986 | 0.996 | 0.984 | 0.987 | 1.000 | 0.999 | 1.000 | 0.970 | 0.998 | 0.999 | 0.999 | 0.971 | 0.998 | 0.988 | 0.990 | 0.990 | 0.977 | 0.987 | 0.986 | 0.990 | 0.988 | 0.996 | 0.996 | 0.972 | 0.987 |
| t-exp | 0.973 | 0.966 | 0.966 | 1.000 | 0.967 | 0.975 | 0.945 | 0.944 | 0.944 | 0.942 | 0.959 | 0.967 | 0.945 | 0.967 | 0.966 | 0.970 | 1.000 | 0.980 | 0.970 | 0.969 | 1.000 | 0.980 | 0.992 | 0.990 | 0.989 | 0.999 | 0.994 | 0.941 | 0.991 | 0.954 | 0.982 | 0.982 | 1.000 | 0.994 |
| lap-hged | 0.999 | 0.997 | 0.997 | 0.980 | 0.997 | 0.999 | 0.986 | 0.985 | 0.985 | 0.984 | 0.995 | 0.989 | 0.985 | 0.998 | 0.997 | 0.998 | 0.980 | 1.000 | 0.999 | 0.999 | 0.981 | 1.000 | 0.995 | 0.996 | 0.996 | 0.985 | 0.994 | 0.984 | 0.996 | 0.989 | 0.999 | 0.999 | 0.982 | 0.994 |
| n-ex | 0.999 | 0.999 | 0.999 | 0.970 | 0.998 | 0.999 | 0.992 | 0.991 | 0.991 | 0.991 | 0.999 | 0.990 | 0.992 | 0.999 | 0.999 | 0.999 | 0.970 | 0.999 | 1.000 | 1.000 | 0.971 | 0.998 | 0.990 | 0.992 | 0.992 | 0.977 | 0.988 | 0.991 | 0.992 | 0.993 | 0.998 | 0.998 | 0.972 | 0.988 |
| lap-hgt | 0.999 | 0.999 | 0.999 | 0.969 | 0.998 | 0.999 | 0.992 | 0.991 | 0.991 | 0.991 | 0.999 | 0.990 | 0.992 | 0.999 | 0.999 | 0.999 | 0.969 | 0.999 | 1.000 | 1.000 | 0.971 | 0.998 | 0.990 | 0.992 | 0.992 | 0.976 | 0.988 | 0.991 | 0.991 | 0.993 | 0.998 | 0.998 | 0.972 | 0.988 |
| t-hged | 0.974 | 0.967 | 0.967 | 1.000 | 0.969 | 0.976 | 0.947 | 0.946 | 0.946 | 0.944 | 0.961 | 0.969 | 0.946 | 0.968 | 0.967 | 0.971 | 1.000 | 0.981 | 0.971 | 0.971 | 1.000 | 0.981 | 0.993 | 0.991 | 0.990 | 0.999 | 0.995 | 0.942 | 0.992 | 0.956 | 0.984 | 0.983 | 1.000 | 0.995 |
| t-hgt | 0.998 | 0.997 | 0.997 | 0.980 | 0.997 | 0.998 | 0.985 | 0.984 | 0.984 | 0.983 | 0.995 | 0.988 | 0.984 | 0.997 | 0.996 | 0.998 | 0.980 | 1.000 | 0.998 | 0.998 | 0.981 | 1.000 | 0.995 | 0.996 | 0.996 | 0.986 | 0.995 | 0.983 | 0.996 | 0.987 | 0.999 | 0.999 | 0.982 | 0.994 |
| n-hged | 0.990 | 0.986 | 0.986 | 0.992 | 0.986 | 0.991 | 0.974 | 0.973 | 0.973 | 0.972 | 0.984 | 0.988 | 0.973 | 0.987 | 0.986 | 0.988 | 0.992 | 0.995 | 0.990 | 0.990 | 0.993 | 0.995 | 1.000 | 1.000 | 1.000 | 0.995 | 1.000 | 0.970 | 1.000 | 0.980 | 0.997 | 0.997 | 0.993 | 0.999 |
| n-hgt | 0.992 | 0.988 | 0.988 | 0.989 | 0.988 | 0.992 | 0.977 | 0.977 | 0.977 | 0.975 | 0.986 | 0.990 | 0.977 | 0.989 | 0.988 | 0.990 | 0.990 | 0.996 | 0.992 | 0.992 | 0.991 | 0.996 | 1.000 | 1.000 | 1.000 | 0.994 | 0.999 | 0.974 | 1.000 | 0.983 | 0.998 | 0.998 | 0.991 | 0.999 |
| gt-ht | 0.992 | 0.988 | 0.988 | 0.989 | 0.988 | 0.993 | 0.978 | 0.977 | 0.977 | 0.976 | 0.987 | 0.991 | 0.977 | 0.989 | 0.988 | 0.990 | 0.989 | 0.996 | 0.992 | 0.992 | 0.990 | 0.996 | 1.000 | 1.000 | 1.000 | 0.994 | 0.999 | 0.974 | 1.000 | 0.983 | 0.998 | 0.998 | 0.991 | 0.999 |
| ged-ht | 0.979 | 0.972 | 0.973 | 0.998 | 0.974 | 0.981 | 0.954 | 0.953 | 0.953 | 0.951 | 0.967 | 0.974 | 0.953 | 0.974 | 0.972 | 0.977 | 0.999 | 0.985 | 0.977 | 0.976 | 0.999 | 0.986 | 0.995 | 0.994 | 0.994 | 1.000 | 0.997 | 0.949 | 0.995 | 0.962 | 0.988 | 0.988 | 0.999 | 0.997 |
| lap-ht | 0.989 | 0.984 | 0.985 | 0.994 | 0.985 | 0.990 | 0.970 | 0.969 | 0.969 | 0.968 | 0.981 | 0.985 | 0.970 | 0.985 | 0.984 | 0.987 | 0.994 | 0.994 | 0.988 | 0.988 | 0.995 | 0.995 | 1.000 | 0.999 | 0.999 | 0.997 | 1.000 | 0.966 | 0.999 | 0.976 | 0.996 | 0.996 | 0.995 | 1.000 |
| t-ht | 0.986 | 0.990 | 0.990 | 0.940 | 0.986 | 0.985 | 1.000 | 1.000 | 1.000 | 1.000 | 0.995 | 0.989 | 1.000 | 0.990 | 0.990 | 0.986 | 0.941 | 0.984 | 0.991 | 0.991 | 0.942 | 0.983 | 0.970 | 0.974 | 0.974 | 0.949 | 0.966 | 1.000 | 0.971 | 0.998 | 0.983 | 0.984 | 0.943 | 0.965 |
| n-ht | 0.992 | 0.988 | 0.988 | 0.991 | 0.988 | 0.993 | 0.975 | 0.974 | 0.974 | 0.973 | 0.985 | 0.988 | 0.974 | 0.989 | 0.988 | 0.990 | 0.991 | 0.996 | 0.992 | 0.991 | 0.992 | 0.996 | 1.000 | 1.000 | 1.000 | 0.995 | 0.999 | 0.971 | 1.000 | 0.981 | 0.998 | 0.998 | 0.992 | 0.999 |
| gt-hn | 0.989 | 0.991 | 0.991 | 0.954 | 0.988 | 0.989 | 0.999 | 0.999 | 0.999 | 0.999 | 0.995 | 0.996 | 0.999 | 0.991 | 0.991 | 0.988 | 0.954 | 0.989 | 0.993 | 0.993 | 0.956 | 0.987 | 0.980 | 0.983 | 0.983 | 0.962 | 0.976 | 0.998 | 0.981 | 1.000 | 0.989 | 0.990 | 0.957 | 0.975 |
| ged-hn | 0.997 | 0.995 | 0.995 | 0.982 | 0.995 | 0.997 | 0.986 | 0.985 | 0.985 | 0.984 | 0.994 | 0.992 | 0.985 | 0.996 | 0.995 | 0.996 | 0.982 | 0.999 | 0.998 | 0.998 | 0.984 | 0.999 | 0.997 | 0.998 | 0.998 | 0.988 | 0.996 | 0.983 | 0.998 | 0.989 | 1.000 | 1.000 | 0.984 | 0.996 |
| t-hn | 0.997 | 0.995 | 0.996 | 0.982 | 0.995 | 0.998 | 0.986 | 0.985 | 0.985 | 0.984 | 0.994 | 0.992 | 0.986 | 0.996 | 0.995 | 0.996 | 0.982 | 0.999 | 0.998 | 0.998 | 0.983 | 0.999 | 0.997 | 0.998 | 0.998 | 0.988 | 0.996 | 0.984 | 0.998 | 0.990 | 1.000 | 1.000 | 0.984 | 0.996 |
| lap-hn | 0.975 | 0.968 | 0.968 | 1.000 | 0.970 | 0.977 | 0.948 | 0.947 | 0.947 | 0.945 | 0.962 | 0.969 | 0.947 | 0.969 | 0.968 | 0.972 | 1.000 | 0.982 | 0.972 | 0.972 | 1.000 | 0.982 | 0.993 | 0.991 | 0.991 | 0.999 | 0.995 | 0.943 | 0.992 | 0.957 | 0.984 | 0.984 | 1.000 | 0.996 |
| n-hn | 0.989 | 0.984 | 0.985 | 0.994 | 0.985 | 0.990 | 0.969 | 0.968 | 0.968 | 0.967 | 0.981 | 0.984 | 0.968 | 0.985 | 0.984 | 0.987 | 0.994 | 0.994 | 0.988 | 0.988 | 0.995 | 0.994 | 0.999 | 0.999 | 0.999 | 0.997 | 1.000 | 0.965 | 0.999 | 0.975 | 0.996 | 0.996 | 0.996 | 1.000 |

Table A3. Spanish dairy farms: posterior means of model parameters

| Model label | $\beta_0$ | $\beta_1$ | $\beta_2$ | $\beta_3$ | $\beta_4$ | $\beta_5$ | $\beta_{2,2}$ | $\beta_{2,3}$ | $\beta_{2,4}$ | $\beta_{2,5}$ | $\beta_{3,3}$ | $\beta_{3,4}$ | $\beta_{3,5}$ | $\beta_{4,4}$ | $\beta_{4,5}$ | $\beta_{5,5}$ | $\sigma_u$ | $\sigma_v$ | $\tau$ | $\psi_u$ | $\psi_v$ | $\nu_u$ | $\nu_v$ |
|---|---|---|---|---|---|---|---|---|---|---|---|---|---|---|---|---|---|---|---|---|---|---|---|
| GT-GB2 | 0.046 | 0.006 | 0.609 | 0.028 | 0.019 | 0.439 | 0.175 | -0.036 | 0.196 | -0.230 | -0.071 | 0.008 | 0.044 | -0.063 | -0.055 | 0.075 | 0.271 | 0.955 | 0.299 | 21.774 | 3.225 | 31.589 | 9.323 |
| LAP-HN | 0.124 | 0.006 | 0.607 | 0.030 | 0.020 | 0.439 | 0.172 | -0.036 | 0.194 | -0.225 | -0.077 | 0.010 | 0.045 | -0.059 | -0.052 | 0.073 | 0.179 | 0.617 | (1) | (2) | (1) | (inf) | (inf) |
| LAP-HT | 0.122 | 0.005 | 0.605 | 0.031 | 0.020 | 0.440 | 0.164 | -0.034 | 0.194 | -0.216 | -0.079 | 0.012 | 0.043 | -0.059 | -0.053 | 0.071 | 0.173 | 0.624 | (1) | (2) | (1) | 48.874 | (inf) |
| GT-N | 0.123 | 0.005 | 0.610 | 0.028 | 0.020 | 0.438 | 0.172 | -0.040 | 0.200 | -0.224 | -0.074 | 0.010 | 0.045 | -0.059 | -0.056 | 0.073 | 0.179 | 0.654 | (1) | (2) | 1.645 | (inf) | 21.330 |
| GT-GG | 0.048 | 0.006 | 0.607 | 0.028 | 0.019 | 0.440 | 0.174 | -0.038 | 0.190 | -0.225 | -0.070 | 0.011 | 0.042 | -0.059 | -0.051 | 0.074 | 0.284 | 0.944 | 0.329 | 24.538 | 2.946 | (inf) | 8.537 |
| GT-HT | 0.117 | 0.006 | 0.606 | 0.029 | 0.020 | 0.440 | 0.174 | -0.044 | 0.198 | -0.226 | -0.072 | 0.011 | 0.046 | -0.057 | -0.057 | 0.073 | 0.168 | 0.688 | (1) | (2) | 1.754 | 44.435 | 2.122 |
| T-HT | 0.115 | 0.006 | 0.608 | 0.027 | 0.020 | 0.439 | 0.178 | -0.041 | 0.199 | -0.232 | -0.071 | 0.008 | 0.044 | -0.056 | -0.057 | 0.076 | 0.166 | 0.724 | (1) | (2) | (2) | 41.358 | 6.253 |
| T-HN | 0.120 | 0.006 | 0.610 | 0.027 | 0.020 | 0.439 | 0.179 | -0.044 | 0.204 | -0.234 | -0.069 | 0.007 | 0.045 | -0.057 | -0.058 | 0.076 | 0.178 | 0.688 | (1) | (2) | (2) | (inf) | 5.491 |
| GED-HN | 0.124 | 0.006 | 0.607 | 0.030 | 0.020 | 0.439 | 0.170 | -0.038 | 0.192 | -0.223 | -0.076 | 0.011 | 0.045 | -0.058 | -0.051 | 0.072 | 0.179 | 0.630 | (1) | (2) | 1.339 | (inf) | (inf) |
| TLAP-GB2 | 0.121 | 0.005 | 0.609 | 0.029 | 0.022 | 0.438 | 0.171 | -0.039 | 0.189 | -0.222 | -0.077 | 0.012 | 0.046 | -0.057 | -0.051 | 0.071 | 0.235 | 0.593 | 0.856 | 4.897 | (1) | 26.743 | 43.812 |
| T-GB2 | 0.052 | 0.006 | 0.601 | 0.030 | 0.020 | 0.442 | 0.170 | -0.042 | 0.194 | -0.220 | -0.068 | 0.010 | 0.042 | -0.054 | -0.057 | 0.073 | 0.262 | 0.917 | 0.335 | 22.276 | (2) | 31.473 | 12.138 |
| GED-HT | 0.120 | 0.006 | 0.606 | 0.030 | 0.020 | 0.439 | 0.173 | -0.038 | 0.201 | -0.226 | -0.077 | 0.009 | 0.045 | -0.060 | -0.057 | 0.074 | 0.170 | 0.656 | (1) | (2) | 1.826 | 44.537 | (inf) |
| LAP-GG | 0.115 | 0.005 | 0.606 | 0.030 | 0.021 | 0.440 | 0.163 | -0.033 | 0.198 | -0.217 | -0.078 | 0.010 | 0.043 | -0.055 | -0.055 | 0.071 | 0.193 | 0.637 | 0.853 | 2.321 | (1) | (inf) | (inf) |
| GT-HGT | 0.125 | 0.005 | 0.609 | 0.028 | 0.020 | 0.439 | 0.178 | -0.041 | 0.199 | -0.232 | -0.074 | 0.009 | 0.046 | -0.058 | -0.057 | 0.075 | 0.184 | 0.644 | (1) | 2.646 | 1.562 | 31.113 | 23.184 |
| LAP-HGT | 0.128 | 0.005 | 0.608 | 0.029 | 0.021 | 0.439 | 0.178 | -0.040 | 0.193 | -0.231 | -0.075 | 0.010 | 0.046 | -0.061 | -0.052 | 0.074 | 0.188 | 0.614 | (1) | 2.764 | (1) | 3.978 | (inf) |
| GED-GB2 | 0.102 | 0.006 | 0.607 | 0.028 | 0.021 | 0.440 | 0.173 | -0.043 | 0.198 | -0.222 | -0.073 | 0.008 | 0.044 | -0.059 | -0.056 | 0.072 | 0.288 | 0.717 | 0.714 | 8.347 | 1.160 | 25.795 | (inf) |
| GED-GG | 0.099 | 0.006 | 0.604 | 0.030 | 0.021 | 0.440 | 0.168 | -0.039 | 0.196 | -0.221 | -0.075 | 0.009 | 0.045 | -0.053 | -0.057 | 0.072 | 0.232 | 0.729 | 0.744 | 4.813 | 1.186 | (inf) | (inf) |
| T-HGT | 0.119 | 0.006 | 0.608 | 0.027 | 0.020 | 0.440 | 0.180 | -0.047 | 0.200 | -0.232 | -0.070 | 0.007 | 0.047 | -0.056 | -0.057 | 0.075 | 0.172 | 0.783 | (1) | 2.394 | (2) | 33.789 | 6.124 |
| GED-HGT | 0.127 | 0.006 | 0.608 | 0.029 | 0.021 | 0.440 | 0.178 | -0.043 | 0.194 | -0.231 | -0.075 | 0.009 | 0.047 | -0.060 | -0.052 | 0.074 | 0.185 | 0.618 | (1) | 2.738 | 1.200 | 29.415 | (inf) |
| LAP-HGED | 0.125 | 0.005 | 0.609 | 0.029 | 0.020 | 0.439 | 0.167 | -0.039 | 0.193 | -0.218 | -0.075 | 0.011 | 0.045 | -0.056 | -0.052 | 0.070 | 0.183 | 0.619 | (1) | 2.116 | (1) | (inf) | (inf) |
| GT-HGED | 0.116 | 0.006 | 0.609 | 0.028 | 0.020 | 0.439 | 0.173 | -0.040 | 0.200 | -0.228 | -0.074 | 0.008 | 0.046 | -0.058 | -0.056 | 0.074 | 0.173 | 0.680 | (1) | 1.952 | 1.761 | (inf) | 2.743 |
| T-EXP | 0.074 | 0.006 | 0.601 | 0.031 | 0.019 | 0.441 | 0.179 | -0.048 | 0.205 | -0.227 | -0.069 | 0.007 | 0.045 | -0.059 | -0.060 | 0.074 | 0.957 | 0.913 | (1) | (2) | (2) | (inf) | 11.182 |
| T-HGED | 0.112 | 0.006 | 0.608 | 0.027 | 0.019 | 0.440 | 0.179 | -0.043 | 0.198 | -0.232 | -0.069 | 0.010 | 0.044 | -0.055 | -0.059 | 0.076 | 0.165 | 0.728 | (1) | 1.835 | (2) | (inf) | 6.354 |
| GT-EXP | 0.073 | 0.006 | 0.604 | 0.028 | 0.019 | 0.441 | 0.177 | -0.043 | 0.199 | -0.228 | -0.068 | 0.007 | 0.044 | -0.063 | -0.056 | 0.074 | 0.949 | 0.942 | (1) | (2) | 2.956 | (inf) | 9.595 |
| GED-HGED | 0.124 | 0.006 | 0.608 | 0.030 | 0.020 | 0.439 | 0.170 | -0.037 | 0.198 | -0.225 | -0.076 | 0.009 | 0.045 | -0.058 | -0.055 | 0.073 | 0.188 | 0.631 | (1) | 2.733 | 1.291 | (inf) | (inf) |
| GED-EXP | 0.075 | 0.006 | 0.598 | 0.032 | 0.020 | 0.443 | 0.176 | -0.051 | 0.204 | -0.222 | -0.068 | 0.011 | 0.045 | -0.058 | -0.062 | 0.073 | 0.962 | 0.942 | (1) | (1) | 1.479 | (inf) | (inf) |
| N-GG | 0.022 | 0.006 | 0.600 | 0.031 | 0.022 | 0.442 | 0.210 | -0.065 | 0.205 | -0.262 | -0.060 | 0.005 | 0.049 | -0.059 | -0.061 | 0.085 | 0.327 | 0.116 | 0.146 | 28.935 | (2) | (inf) | (inf) |
| N-EX | 0.068 | 0.006 | 0.601 | 0.030 | 0.023 | 0.442 | 0.224 | -0.076 | 0.212 | -0.274 | -0.058 | 0.003 | 0.052 | -0.061 | -0.062 | 0.087 | 0.925 | 0.146 | (1) | (1) | (2) | (inf) | (inf) |
| N-HT | 0.098 | 0.006 | 0.603 | 0.028 | 0.023 | 0.442 | 0.234 | -0.083 | 0.215 | -0.286 | -0.053 | 0.000 | 0.054 | -0.056 | -0.062 | 0.091 | 0.148 | 0.969 | (1) | (2) | (2) | 22.867 | (inf) |
| LAP-EXP | 0.088 | 0.006 | 0.592 | 0.036 | 0.020 | 0.443 | 0.114 | -0.021 | 0.198 | -0.156 | -0.084 | 0.024 | 0.037 | -0.058 | -0.063 | 0.053 | 0.145 | 0.725 | (1) | (1) | (1) | (inf) | (inf) |
| N-HN | 0.108 | 0.006 | 0.605 | 0.027 | 0.024 | 0.441 | 0.239 | -0.084 | 0.210 | -0.292 | -0.051 | 0.000 | 0.053 | -0.050 | -0.062 | 0.093 | 0.169 | 0.942 | (1) | (2) | (2) | (inf) | (inf) |
| N-HGT | 0.084 | 0.006 | 0.601 | 0.029 | 0.024 | 0.443 | 0.230 | -0.079 | 0.214 | -0.280 | -0.056 | 0.002 | 0.052 | -0.056 | -0.065 | 0.090 | 0.126 | 0.170 | (1) | 1.539 | (2) | 38.667 | (inf) |
| N-HGED | 0.082 | 0.006 | 0.600 | 0.030 | 0.023 | 0.443 | 0.226 | -0.077 | 0.214 | -0.277 | -0.055 | 0.000 | 0.051 | -0.057 | -0.063 | 0.089 | 0.118 | 0.163 | (1) | 1.294 | (2) | (inf) | (inf) |
| LAP-GAM | 0.085 | 0.006 | 0.592 | 0.037 | 0.019 | 0.443 | 0.112 | -0.018 | 0.199 | -0.156 | -0.085 | 0.026 | 0.035 | -0.059 | -0.063 | 0.054 | 0.164 | 0.731 | 0.952 | (1) | (1) | (inf) | (inf) |
| Average (uniform) | 0.098 | 0.006 | 0.607 | 0.029 | 0.020 | 0.439 | 0.172 | -0.038 | 0.196 | -0.226 | -0.074 | 0.010 | 0.044 | -0.059 | -0.054 | 0.074 | 0.206 | 0.074 | 0.377 | 18.310 | 2.333 | 37.857 | 13.097 |
| Average (OS prior) | 0.120 | 0.006 | 0.607 | 0.030 | 0.020 | 0.439 | 0.171 | -0.037 | 0.195 | -0.225 | -0.076 | 0.010 | 0.045 | -0.059 | -0.053 | 0.073 | 0.181 | 0.065 | 0.471 | 12.751 | 1.743 | 44.511 | 12.732 |

Table A4. Spanish dairy farms: posterior standard deviations of model parameters

| Model label | $\beta_0$ | $\beta_1$ | $\beta_2$ | $\beta_3$ | $\beta_4$ | $\beta_5$ | $\beta_{2,2}$ | $\beta_{2,3}$ | $\beta_{2,4}$ | $\beta_{2,5}$ | $\beta_{3,3}$ | $\beta_{3,4}$ | $\beta_{3,5}$ | $\beta_{4,4}$ | $\beta_{4,5}$ | $\beta_{5,5}$ | $\sigma_u$ | $\sigma_v$ | $\tau$ | $\psi_u$ | $\psi_v$ | $\nu_u$ | $\nu_v$ |
|---|---|---|---|---|---|---|---|---|---|---|---|---|---|---|---|---|---|---|---|---|---|---|---|
| GT-GB2 | 0.030 | 0.002 | 0.022 | 0.011 | 0.013 | 0.012 | 0.067 | 0.061 | 0.065 | 0.072 | 0.024 | 0.041 | 0.030 | 0.041 | 0.037 | 0.022 | 0.057 | 0.013 | 0.218 | 26.442 | 1.181 | 26.577 | 12.253 |
| LAP-HN | 0.011 | 0.002 | 0.022 | 0.011 | 0.012 | 0.012 | 0.066 | 0.056 | 0.064 | 0.072 | 0.023 | 0.037 | 0.029 | 0.041 | 0.036 | 0.022 | 0.008 | 0.004 | | | | | |
| LAP-HT | 0.011 | 0.002 | 0.022 | 0.011 | 0.012 | 0.012 | 0.066 | 0.056 | 0.063 | 0.073 | 0.022 | 0.037 | 0.029 | 0.041 | 0.036 | 0.023 | 0.009 | 0.004 | | | | | 31.856 |
| GT-N | 0.012 | 0.002 | 0.021 | 0.011 | 0.012 | 0.012 | 0.068 | 0.061 | 0.065 | 0.073 | 0.024 | 0.038 | 0.031 | 0.042 | 0.037 | 0.023 | 0.009 | 0.008 | | | | 0.858 | 24.645 |
| GT-GG | 0.032 | 0.002 | 0.021 | 0.011 | 0.012 | 0.011 | 0.070 | 0.061 | 0.065 | 0.074 | 0.024 | 0.040 | 0.031 | 0.043 | 0.036 | 0.023 | 0.068 | 0.014 | 0.260 | 27.440 | 1.096 | | 10.608 |
| GT-HT | 0.013 | 0.002 | 0.022 | 0.011 | 0.013 | 0.012 | 0.066 | 0.059 | 0.065 | 0.073 | 0.024 | 0.038 | 0.030 | 0.041 | 0.036 | 0.023 | 0.012 | 0.009 | | | 0.925 | 32.128 | 23.822 |
| T-HT | 0.013 | 0.002 | 0.021 | 0.011 | 0.012 | 0.012 | 0.068 | 0.061 | 0.065 | 0.073 | 0.024 | 0.039 | 0.030 | 0.042 | 0.037 | 0.022 | 0.012 | 0.007 | | | | 29.747 | 2.411 |
| T-HN | 0.012 | 0.002 | 0.021 | 0.011 | 0.013 | 0.012 | 0.067 | 0.059 | 0.064 | 0.073 | 0.024 | 0.039 | 0.030 | 0.042 | 0.036 | 0.023 | 0.009 | 0.007 | | | | | 1.655 |
| GED-HN | 0.012 | 0.002 | 0.022 | 0.011 | 0.012 | 0.012 | 0.066 | 0.057 | 0.065 | 0.074 | 0.023 | 0.038 | 0.030 | 0.041 | 0.036 | 0.023 | 0.009 | 0.007 | | | | 0.142 | |
| TLAP-GB2 | 0.015 | 0.002 | 0.022 | 0.011 | 0.012 | 0.012 | 0.067 | 0.060 | 0.063 | 0.073 | 0.024 | 0.037 | 0.031 | 0.041 | 0.036 | 0.023 | 0.018 | 0.006 | 0.105 | 6.909 | | 27.509 | 31.646 |
| T-GB2 | 0.029 | 0.002 | 0.021 | 0.011 | 0.012 | 0.012 | 0.068 | 0.060 | 0.064 | 0.075 | 0.025 | 0.038 | 0.030 | 0.044 | 0.037 | 0.023 | 0.057 | 0.012 | 0.222 | 28.984 | | 28.282 | 8.566 |
| GED-HT | 0.012 | 0.002 | 0.021 | 0.011 | 0.011 | 0.012 | 0.068 | 0.057 | 0.064 | 0.075 | 0.024 | 0.037 | 0.029 | 0.041 | 0.036 | 0.023 | 0.011 | 0.008 | | | | 0.152 | 29.843 |
| LAP-GG | 0.019 | 0.002 | 0.022 | 0.011 | 0.012 | 0.012 | 0.066 | 0.056 | 0.064 | 0.073 | 0.023 | 0.037 | 0.029 | 0.041 | 0.036 | 0.022 | 0.021 | 0.006 | 0.127 | 0.502 | | | |
| GT-HGT | 0.016 | 0.002 | 0.022 | 0.011 | 0.012 | 0.012 | 0.070 | 0.058 | 0.064 | 0.076 | 0.024 | 0.038 | 0.030 | 0.044 | 0.036 | 0.024 | 0.023 | 0.010 | | 1.006 | 0.854 | 28.038 | 26.181 |
| LAP-HGT | 0.012 | 0.002 | 0.022 | 0.011 | 0.012 | 0.012 | 0.068 | 0.058 | 0.064 | 0.074 | 0.023 | 0.038 | 0.030 | 0.043 | 0.035 | 0.023 | 0.016 | 0.004 | | 0.918 | | 28.713 | |
| GED-GB2 | 0.030 | 0.002 | 0.022 | 0.011 | 0.013 | 0.012 | 0.065 | 0.057 | 0.064 | 0.072 | 0.025 | 0.038 | 0.030 | 0.043 | 0.036 | 0.023 | 0.037 | 0.014 | 0.218 | 16.847 | 0.245 | 27.417 | |
| GED-GG | 0.032 | 0.002 | 0.022 | 0.011 | 0.013 | 0.012 | 0.066 | 0.057 | 0.064 | 0.074 | 0.025 | 0.036 | 0.029 | 0.042 | 0.037 | 0.024 | 0.046 | 0.014 | 0.249 | 11.487 | 0.247 | | |
| T-HGT | 0.017 | 0.002 | 0.022 | 0.011 | 0.012 | 0.012 | 0.069 | 0.060 | 0.066 | 0.075 | 0.025 | 0.039 | 0.030 | 0.042 | 0.037 | 0.024 | 0.026 | 0.010 | | 0.910 | | 30.264 | 2.559 |
| GED-HGT | 0.017 | 0.002 | 0.022 | 0.011 | 0.012 | 0.012 | 0.067 | 0.060 | 0.064 | 0.073 | 0.023 | 0.038 | 0.031 | 0.041 | 0.035 | 0.023 | 0.026 | 0.011 | | 0.903 | 0.182 | 27.842 | |
| LAP-HGED | 0.012 | 0.002 | 0.021 | 0.011 | 0.012 | 0.012 | 0.067 | 0.058 | 0.062 | 0.072 | 0.023 | 0.038 | 0.030 | 0.041 | 0.035 | 0.023 | 0.016 | 0.004 | | 0.294 | | | |
| GT-HGED | 0.017 | 0.002 | 0.022 | 0.011 | 0.012 | 0.012 | 0.067 | 0.058 | 0.064 | 0.072 | 0.023 | 0.038 | 0.030 | 0.043 | 0.037 | 0.023 | 0.024 | 0.011 | | 0.368 | 0.986 | | 25.027 |
| T-EXP | 0.012 | 0.002 | 0.021 | 0.011 | 0.012 | 0.012 | 0.067 | 0.058 | 0.064 | 0.073 | 0.024 | 0.039 | 0.030 | 0.043 | 0.036 | 0.023 | 0.007 | 0.006 | | | | | 5.895 |
| T-HGED | 0.016 | 0.002 | 0.021 | 0.011 | 0.012 | 0.012 | 0.066 | 0.059 | 0.064 | 0.071 | 0.024 | 0.039 | 0.030 | 0.042 | 0.036 | 0.022 | 0.025 | 0.009 | | 0.367 | | | 2.964 |
| GT-EXP | 0.011 | 0.002 | 0.021 | 0.011 | 0.013 | 0.012 | 0.069 | 0.060 | 0.064 | 0.074 | 0.024 | 0.039 | 0.030 | 0.043 | 0.036 | 0.022 | 0.007 | 0.007 | | | | 1.304 | 12.132 |
| GED-HGED | 0.014 | 0.002 | 0.022 | 0.011 | 0.012 | 0.012 | 0.066 | 0.057 | 0.063 | 0.074 | 0.023 | 0.039 | 0.030 | 0.042 | 0.037 | 0.023 | 0.020 | 0.009 | | 0.362 | 0.172 | | |
| GED-EXP | 0.012 | 0.002 | 0.021 | 0.011 | 0.013 | 0.012 | 0.064 | 0.056 | 0.064 | 0.070 | 0.023 | 0.037 | 0.028 | 0.043 | 0.036 | 0.022 | 0.007 | 0.006 | | | | 0.141 | |
| N-GG | 0.014 | 0.002 | 0.021 | 0.011 | 0.013 | 0.012 | 0.063 | 0.056 | 0.066 | 0.068 | 0.023 | 0.041 | 0.028 | 0.044 | 0.037 | 0.021 | 0.048 | 0.004 | 0.059 | 27.883 | | | |
| N-EX | 0.011 | 0.002 | 0.021 | 0.011 | 0.013 | 0.012 | 0.064 | 0.055 | 0.065 | 0.070 | 0.023 | 0.040 | 0.029 | 0.044 | 0.037 | 0.021 | 0.007 | 0.004 | | | | | |
| N-HT | 0.014 | 0.002 | 0.021 | 0.012 | 0.013 | 0.012 | 0.064 | 0.056 | 0.066 | 0.069 | 0.023 | 0.040 | 0.029 | 0.044 | 0.038 | 0.021 | 0.016 | 0.005 | | | | | 21.905 |
| LAP-EXP | 0.011 | 0.002 | 0.020 | 0.010 | 0.012 | 0.011 | 0.062 | 0.055 | 0.066 | 0.071 | 0.023 | 0.036 | 0.028 | 0.043 | 0.036 | 0.023 | 0.007 | 0.004 | | | | | |
| N-HN | 0.012 | 0.002 | 0.021 | 0.011 | 0.013 | 0.012 | 0.063 | 0.055 | 0.066 | 0.069 | 0.024 | 0.041 | 0.028 | 0.044 | 0.037 | 0.021 | 0.009 | 0.005 | | | | | |
| N-HGT | 0.017 | 0.002 | 0.021 | 0.011 | 0.013 | 0.012 | 0.063 | 0.056 | 0.066 | 0.068 | 0.023 | 0.040 | 0.028 | 0.044 | 0.036 | 0.021 | 0.024 | 0.006 | | 0.486 | | 31.978 | |
| N-HGED | 0.017 | 0.002 | 0.021 | 0.011 | 0.013 | 0.012 | 0.063 | 0.057 | 0.067 | 0.067 | 0.023 | 0.040 | 0.029 | 0.044 | 0.037 | 0.021 | 0.022 | 0.005 | | 0.253 | | | |
| LAP-GAM | 0.012 | 0.002 | 0.020 | 0.010 | 0.012 | 0.012 | 0.063 | 0.057 | 0.063 | 0.072 | 0.024 | 0.035 | 0.029 | 0.042 | 0.036 | 0.023 | 0.007 | 0.004 | 0.045 | | | | |

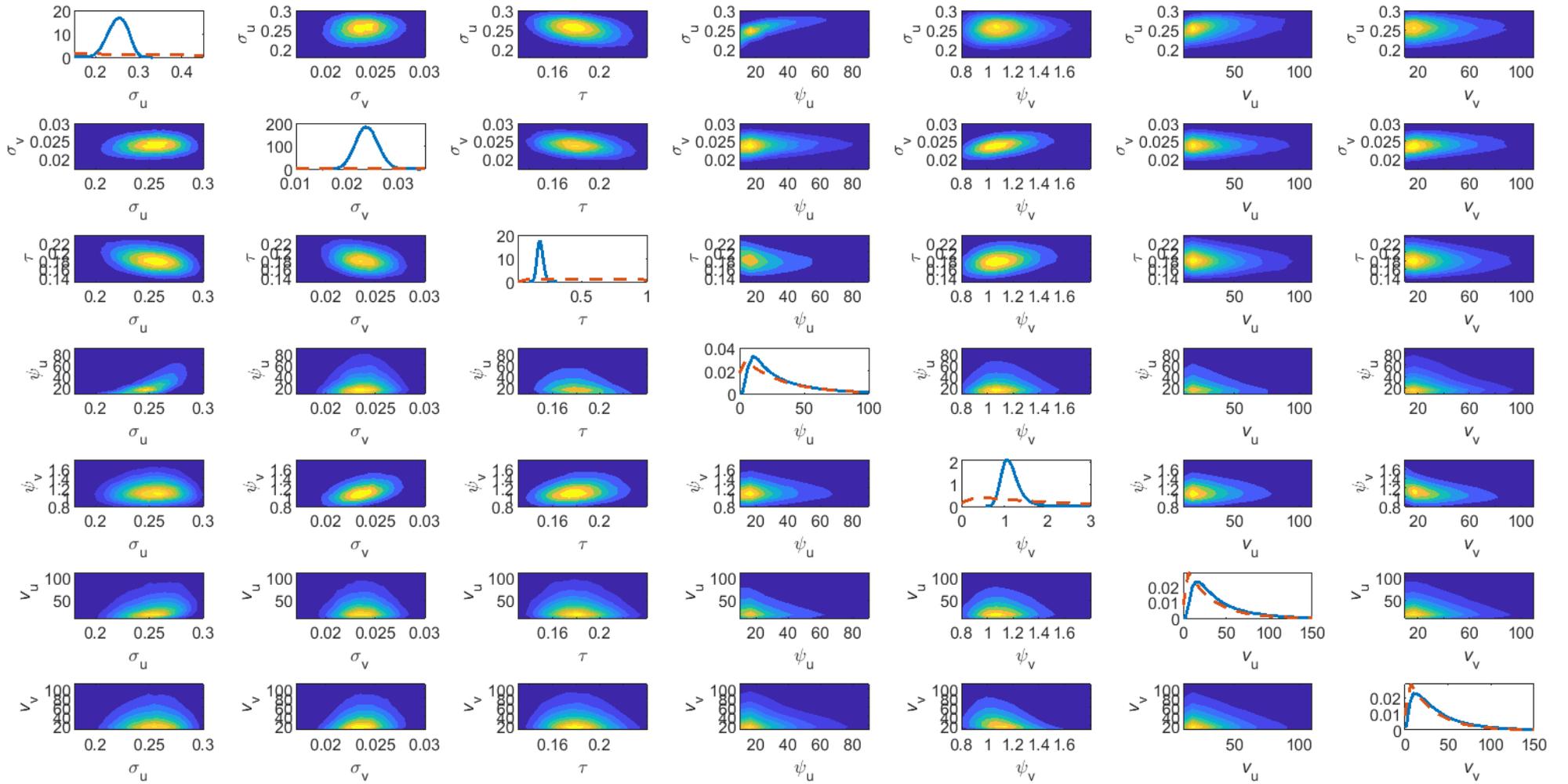

Figure A1. WHO: full heatmap matrix for 2D marginal posterior densities

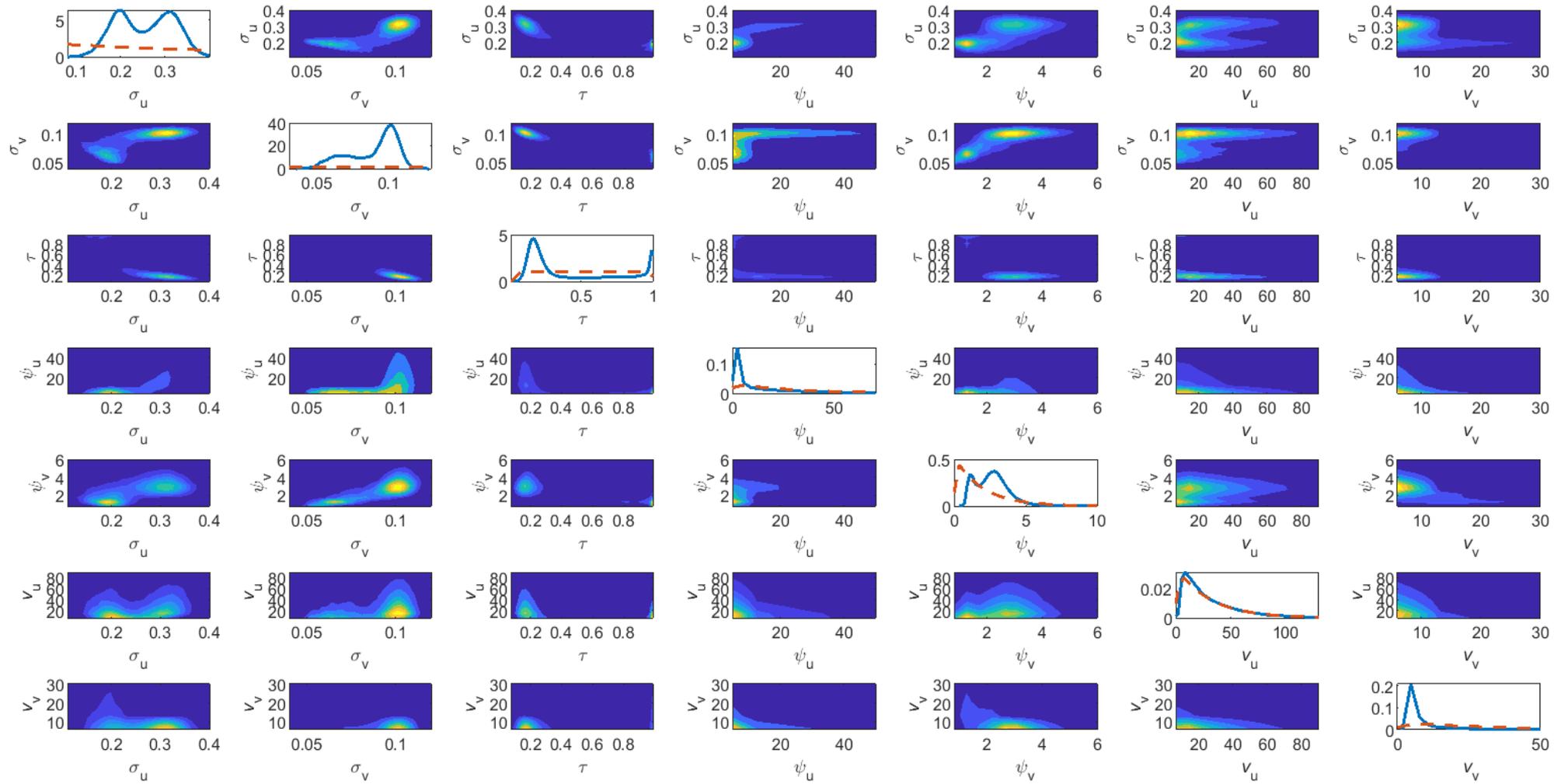

Figure A2. Spanish dairy farms: full heatmap matrix for 2D marginal posterior densities